\documentclass[12pt,letterpaper]{article}
 
\usepackage{amsmath,amssymb}
\usepackage{array}
\usepackage{cite}
\usepackage{enumerate}

\usepackage{float}
\usepackage{amsmath,amsfonts,graphicx,bm}
\usepackage[usenames]{color}
\usepackage{multirow}

\usepackage{amsmath,amsfonts,graphicx,bbm,multicol}

\usepackage{subfigure}
\usepackage{titlesec}
\newcommand*{\justifyheading}{\centering}
\titleformat{\chapter}[display]
  {\normalfont\huge\bfseries\justifyheading}{\chaptertitlename\ \thechapter}
  {20pt}{\Huge}
\titleformat{\section}
  {\normalfont\Large\bfseries\justifyheading}{\thesection}{1em}{}
\titleformat{\subsection}
  {\normalfont\Large\bfseries\justifyheading}{\thesubsection}{1em}{}

\newcolumntype{x}[1]{%
{\centering\hspace{0pt}}p{#1}}%

\def\BR{{\rm BR}}
\newcommand{\be}{\begin{equation}}
\newcommand{\ee}{\end{equation}}
\newcommand{\bea}{\begin{eqnarray}}
\newcommand{\eea}{\end{eqnarray}}

\begin{document}

\baselineskip=18pt

\def\thesection{\Roman{section}} 
\def\thesubsection{\indent\Alph{subsection}}

\newcommand{\eps}{\epsilon}
\newcommand{\pslash}{\!\not\! p}
\newcommand{\I}{\rm 1\kern-.24em l} 
\newcommand{\Tr}{\mathop{\rm Tr}}



\thispagestyle{empty}
\vspace{20pt}
\font\cmss=cmss10 \font\cmsss=cmss10 at 7pt

\begin{flushright}
PITT-PACC-1313
\end{flushright}

\hfill
\vspace{20pt}

\begin{center}
{\Large \textbf
{
Higgs Bosons from Top Quark Decays
}}

\vspace{5pt}

{\large  
Tao Han\footnote{than@pitt.edu}$^{,a,b,c}$ and 
Richard Ruiz\footnote{rer50@pitt.edu}$^{,a}$
}

\vspace{10pt}
\textit{$^{a}$Pittsburgh Particle physics, Astronomy, and Cosmology Center\\
Department of Physics $\&$ Astronomy, University of Pittsburgh, Pittsburgh, PA 15260, USA\\
$^{b}$
Center for High Energy Physics, Tsinghua University, Beijing 100084, China\\
$^{c}$
Korea Institute for Advanced Study (KIAS), Seoul 130-012, Korea
}
\vspace{10pt}
\end{center}

\vspace{15pt}

\begin{abstract}
In light of the discovery of a Standard Model (SM)-like Higgs boson ($h$) at the LHC, 
we investigate the top quark to Higgs boson transition $t\rightarrow W^{*}bh$, which is the leading $t\to h$ decay mode in the SM. 
We find the decay branching fraction to be $1.80\times 10^{-9}$.
In comparison, the two-body, loop-induced $t\rightarrow ch$ transition occurs at $\sim10^{-14}$ in the SM.
We consider the consequences of gauge invariant dimension-6 operators affecting the $t\bar{t}h$ interaction
and find that the decay branching fraction may be increased by a factor of two within current constraints on the coupling parameters from collider experiments. 
We also extend the calculation to the CP-conserving Type I and Type II Two Higgs Doublet Models (2HDM), including both CP-even and CP-odd Higgs bosons. 
For neutral scalar masses at about $100$ GeV, the decay rates can be several times larger than the SM result in the allowed range of model parameters.
Observation prospects at present and future colliders are briefly addressed.
\end{abstract}


\vfill\eject
\noindent


\section{INTRODUCTION}
\label{intro.SEC}

The discovery of a light, Standard Model (SM)-like Higgs boson at the CERN Large Hadron Collider (LHC)~\cite{:2012gk} 
is a tremendous step towards understanding the  underlying mechanism of electroweak (EW) symmetry breaking (EWSB). 
The observed signal, consistent with the leading production mechanism $gg \to h$,
indicates the existence of the Higgs boson coupling to the top-quark~\cite{Chatrchyan:2013lba}. 
Ultimately, the $t \bar t h$ coupling may be determined at the LHC luminosity upgrade and at a high energy $e^{+}e^{-}$ linear collider~\cite{Dawson:2013bba}.
Regardless of their rarity, a Higgs boson that is less massive than the top quark implies that $t\to h$ transitions exist.
With an annual luminosity of $\mathcal{L}=100~\text{fb}^{-1}/\text{yr}$, the 14 TeV LHC will produce over 90 million $t\overline{t}$ pairs a year~\cite{Czakon:2013goa}.
Thus, searches for $t\to h$ transitions that are sensitive to new physics scenarios are an essential part of the LHC program.
For example: the rare decay involving the Flavor Changing Neutral Current (FCNC) 
\be
t\rightarrow ch.
\label{eq:ch}
\ee
This process is particularly interesting for several reasons. 
At leading order, it is induced at one-loop in the SM and,
due to GIM suppression \cite{Glashow:1970gm,Eilam:1990zc,Mele:1998ag}, its branching fraction is very small, about  $10^{-14}$.
NLO QCD contributions increase this by 10\% \cite{Zhang:2013xya}.
However, new physics beyond the SM (BSM), such as an extended Higgs sector~\cite{Eilam:1990zc,Hou:1991un,Atwood:1996vj,AguilarSaavedra:2004wm,Chen:2013qta} or Supersymmetry (SUSY) \cite{Yang:1997dk,Cao:2007dk,Eilam:2001dh}, can significantly enhance this decay, making it a very sensitive channel to new physics.

\begin{figure}[tb]
\centering
{
	\includegraphics[width=0.9\textwidth]{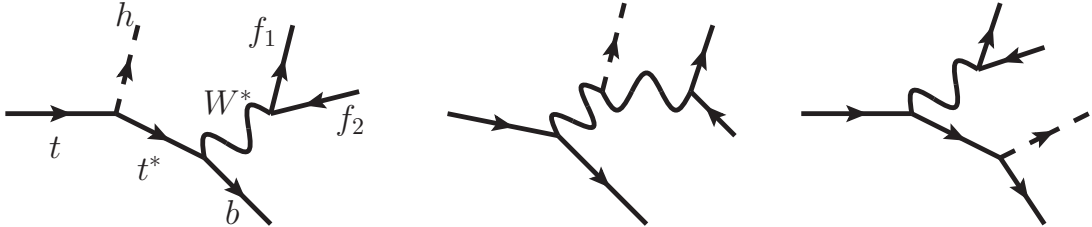}	
}
\caption{Feynman diagrams representing the leading transition $t\to H$ in Eqs.~(\ref{tWbh.EQ}) and (\ref{tWbH.EQ}).
Drawn using the package JaxoDraw~\cite{Binosi:2003yf}.
}
\label{feynman.FIG}
\end{figure}

In this study, we consider another $t\to h$ transition:
\begin{equation}
 t\rightarrow W^{*}b\ h,
\label{tWbh.EQ}
\end{equation}
where the off-shell $W^{*}$ decays to a pair of light fermions. We now know that this is kinematically allowed in the SM. 
Proceeding at tree-level through the diagrams depicted in Fig.~\ref{feynman.FIG}, Eq.~(\ref{tWbh.EQ}) has been previously evaluated \cite{Rizzo:1986sh,Barger:1987bv,Barger:1989ur,Mahlon:1994us,Decker:1991cz,Decker:1992wz,Iltan:2002am}. 
Both the $t\overline{t}h$ and $WWh$ interactions are simultaneously involved, resulting in a certain subtle, but accidental, cancellation. The predicted branching fraction in the SM is about $10^{-9}$. 
Though still small, the rate is significantly larger than that of Eq.~(\ref{eq:ch}), thereby representing the leading $t\to h$ transition in the SM. 
Subsequently, we are  motivated to investigate how sensitive Eq.~(\ref{tWbh.EQ}) is to new physics.

To systematically quantify this sensitivity in a model-independent fashion, we first employ the approach of  Effective Field Theory (EFT). 
In particular, we consider the effects of gauge invariant, dimension-six operators that can alter the $t\overline{t}h$ interaction 
and take into account constraints on anomalous $t\overline{t}h$ couplings imposed by data.

It is highly probable that the scalar sector responsible for the EWSB extends well beyond a solitary Higgs boson. 
For example: in the Two Higgs Doublet Model (2HDM), one of the best motivated SM extensions, an additional scalar $SU(2)_{L}$ doublet is introduced to facilitate EWSB. 
We extend our study into leading $t\rightarrow h$ transitions by considering CP-conserving variants of the so-called Type I and Type II 2HDM,
denoted by 2HDM(I) and 2HDM(II), respectively.
The corresponding decay channel is 
\begin{equation}
 t\rightarrow W^{*}bH\rightarrow 
 f_{1} \bar f_{2}\  bH,
\label{tWbH.EQ}
\end{equation}
where $H$ is generically either one of the two CP-even ($h,~H$) or the CP-odd ($A$) Higgs bosons, and $f_{1}, f_{2}$ are the light fermions in the SM. 
For $h/H$, Eq.~(\ref{tWbH.EQ}) proceeds identically though Fig.~\ref{feynman.FIG}.
For $A$, the middle diagram is absent.

The remainder of this analysis proceeds as follows: 
In section~\ref{th.SEC}, we introduce our theoretical framework and comment on current experimental constraints for each new physics scenario.
We then present in section~\ref{br.SEC} the SM, EFT, 2HDM(I), and 2HDM(II) predictions for the top quark branching fraction of 
Eq.~(\ref{tWbH.EQ}) over respective parameter spaces.
Observation prospects at present and future colliders are briefly addressed in section~\ref{LHC.SEC}.
Finally in section~\ref{conc.SEC}, we summarize our results and conclude.

\section{THEORETICAL FRAMEWORK}
\label{th.SEC}
The theoretical frameworks under consideration include the effective field theory (EFT) for $t\overline{t} h$ interactions up to dimension-six operators (Sec.~\!\!\!\!\!\!\!\!\!\!\ref{eft.SEC}), the two Higgs doublet model of Type I [2HDM(I)] (Sec.~\!\!\!\!\!\!\!\!\!\!\ref{2hdmI.SEC}), and Type II [2HDM(II)] (Sec.~\!\!\!\!\!\!\!\!\!\!\ref{2hdmII.SEC}). Current experimental constraints on the model parameters are also presented. 

\subsection{The SM as an Effective Field Theory}
\label{eft.SEC}
To systematically search for new physics beyond the reach of present-day experiments,
we employ Effective Field Theory (EFT) to model new physical phenomena and linearly realize the SM gauge symmetries\cite{Burges:1983zg,Whisnant:1997qu,Grzadkowski:2010es}. 
After integrating out heavy degrees of freedom at a scale $\Lambda$, the low energy effects can be parameterized by 
\begin{equation}
 \mathcal{L} = \mathcal{L}_{\rm SM}+\mathcal{L}_{\rm Eff.},\quad \mathcal{L}_{\rm Eff.}=\sum_{i,j}\frac{f_{i,j}}{\Lambda^{i}}\mathcal{O}_{i,j},
 \label{formalEffLag.EQ}
\end{equation}
where $\mathcal{L}_{\rm SM}$ is the SM Lagrangian, the $f_{i,j}$ are real, dimensionless ``anomalous couplings'' naturally of order $1\sim 4\pi$, 
and $\mathcal{O}_{i,j}$ represent $SU(3)_{c}\times SU(2)_{L}\times U(1)_{Y}$-invariant, dimension-$(4+i)$ operators constructed solely from SM fields. 
When $f_{i,j}\rightarrow 4\pi$, however, one is likely in the strong coupling regime
and the EFT approach breaks down. Here, $f_{i,j}$ is assumed to be $\mathcal{O}(1)$. 
For the remainder of the text, we consider only the next-to-leading interactions at dimension-six and drop the $i=2$ subscript.

\subsubsection{EFT framework and parameters}
Many linearly independent dimension-six operators can affect the $t\overline{t}h$, $WWh$, $b\overline{b}h$, $tWb$, $htc/u$, or 4-point $tWbh$ vertices
~\cite{Burges:1983zg,Whisnant:1997qu,Gounaris:1996yp,Yang:1997iv,Han:1999xd,DeRujula:1991se,AguilarSaavedra:2008zc,
AguilarSaavedra:2009mx,Grzadkowski:2010es,Einhorn:2013kja,Einhorn:2013tja}.
Results from the ATLAS and CMS experiments indicate that the $WWh$ coupling is close to its SM value~\cite{:2012gk,Chatrchyan:2013lba},
and evidence suggest that the $b\overline{b}h$ coupling cannot be much larger than the SM prediction~\cite{Chatrchyan:2013zna,ATLASHbb}.
As dimension-six $tWbh$ verticies originate from terms of the form $tWb(v+h)$~\cite{Whisnant:1997qu,Yang:1997iv},
the size of anomalous 4-point $tWbh$ couplings are restricted to be small by 
the stringent limits on anomalous $tWb$ couplings~\cite{Abazov:2010jn,Abazov:2012uga,Aaltonen:2012lua,Aad:2012ky,Gounaris:1996yp,Han:1999xd}. 
Anomalous $htc/u$ couplings are constrained to be small~\cite{Chen:2013qta,Craig:2012vj,Atwood:2013ica,TheATLAScollaboration:2013nia}.

As we are interesting in the next-to-leading contribution to the $t\rightarrow W^{*}bh$ transition, 
we consider those operators affecting the weakly constrained $t\overline{t}h$ vertex only.
In the basis of Ref.~\cite{Whisnant:1997qu}, 
the most general $t\overline{t}h$ interaction Lagrangian one can construct using linearly independent dimension-six operators 
requires only two operators~\cite{AguilarSaavedra:2009mx} (one CP-even and one CP-odd):
\begin{eqnarray}
 \mathcal{O}_{t1}=\left(\Phi^\dagger\Phi -\frac{v^{2}}{2}\right)\left(\overline{q_{L}}t_{R}\tilde{\Phi}+\tilde{\Phi}^\dagger\overline{t_{R}}q_{L}\right),
 \quad
 \overline{\mathcal{O}}_{t1}=i\left(\Phi^\dagger\Phi -\frac{v^{2}}{2}\right)\left(\overline{q_{L}}t_{R}\tilde{\Phi}-\tilde{\Phi}^\dagger\overline{t_{R}}q_{L}\right),
 \label{GIOp1.EQ}
 \end{eqnarray}
 where $\Phi$ is the SM Higgs $SU(2)_{L}$ doublet with $U(1)_{Y}$ hypercharge $+1$, 
\begin{equation}
 v=\sqrt{2}\langle\Phi\rangle\approx246~\text{GeV},\quad
 \overline{q_{L}}=(\overline{t_L},\overline{b_{L}}),\quad
 \tilde{\Phi}=i\sigma_{2}\Phi^{*},\quad
 t_{L/R}=P_{L/R}t,
\end{equation}
and $P_{L/R}=\frac{1}{2}(1\mp\gamma^{5})$ is the left/right-handed (LH/RH) chiral projection operator. 
These respectively lead to anomalous scalar- and pseudoscalar-type interactions and 
correspond to the operator $Q_{u\varphi}$ in Refs.~\cite{Grzadkowski:2010es,Einhorn:2013kja}, which assume \textit{complex} Wilson coefficients.
To investigate the sensitivity of operators that select out different kinematic features from those listed above,
we consider also the two redundant\footnote{
Using integration by parts and the appropriate equations of motion, e.g.,
$i\overrightarrow{\not\!\! D} q_{L} = y_{u} u_{R} \tilde{\Phi} + y_{d} d_{R} \Phi$, 
one finds that the operator $\overline{\mathcal{O}}_{t2}$ is linearly dependent on $\mathcal{O}_{t1}$ and $\overline{\mathcal{O}}_{t1}$ plus the bottom quark analogues. Similarly, $\overline{\mathcal{O}}_{\Phi q}^{(1)}$ is linearly dependent on $\mathcal{O}_{t1}$ and $\overline{\mathcal{O}}_{t1}$ \cite{Grzadkowski:2010es}.} (CP-odd) operators
\begin{eqnarray}
  \overline{\mathcal{O}}_{\Phi q}^{(1)}=\left[\Phi^\dagger(D_{\mu}\Phi)+(D_{\mu}\Phi)^\dagger\Phi\right](\overline{q_{L}}\gamma^{\mu}q_{L}),
   \quad
  \overline{\mathcal{O}}_{t2}=\left[\Phi^\dagger(D_{\mu}\Phi)+(D_{\mu}\Phi)^\dagger\Phi\right](\overline{t_{R}}\gamma^{\mu}t_{R}), 
 \label{GIOp2.EQ}
\end{eqnarray}
which respectively lead to anomalous left/right-handed (LH/RH) chiral couplings. 
We do not consider other operators that can affect the $t\rightarrow W^{*}bh$ decay because their Wilson coefficients are strongly constrained by data.

After EWSB, the $t\overline{t}h$ interaction Lagrangian contains four\footnote{
The anomalous LH chiral $b\overline{b}h$ coupling from $\overline{\mathcal{O}}_{\Phi q}^{(1)}$ 
is ignored as its contribution suffers from kinematic and helicity suppression. 
See the discussion in Sec.~\ref{br.SEC}\!\!\!\!\ref{EffBR.SEC}.
} 
new independent terms:
\begin{equation}
 \mathcal{L}_{tth}= - \frac{1}{\sqrt{2}}\overline{t}\left(y_{t}-g^{S}-ig^{P}\gamma^{5}\right)th + \left(\frac{\partial_{\mu}h}{v}\right)\overline{t}\gamma^{\mu}\left(g^{L}P_{L}+g^{R}P_{R}\right)t, 
 \label{eftLag.EQ}
\end{equation}
where $y_{t}$ is the SM top quark Yukawa coupling,
\begin{equation}
y_{t}  = \frac{gm_t}{ \sqrt 2 M_W} \simeq 1,
\end{equation}
and  the anomalous couplings $g^{X}$ beyond the SM (BSM) are
\begin{equation}
 g^{S}=f_{t1}\frac{v^{2}}{\Lambda^{2}},\quad
 g^{P}=\overline{f}_{t1}\frac{v^{2}}{\Lambda^{2}},\quad
 g^{L}=\overline{f}_{\Phi q}^{(1)}\frac{v^{2}}{\Lambda^{2}},\quad
 g^{R}=\overline{f}_{t2}\frac{v^{2}}{\Lambda^{2}}.
 \label{eftCoup.EQ}
\end{equation}
The relative minus signs between $y_{t}$ and $g^{X}$ are arbitrary due to the unknown couplings $f$. 
To better understand the influence of $g^{S}$ and $g^{P}$ on Eq.~(\ref{tWbh.EQ}), it is useful to rewrite the relevant parts of Eq.~(\ref{eftLag.EQ}) as
\begin{equation}
y_{t}-g^{S}-ig^{P}\gamma^{5} =  g^{\rm Eff.}\left( e^{-i\delta_{\rm CP}}P_{R} + e^{i\delta_{\rm CP}} P_{L}\right),
  \label{cpvLag.EQ} 
\end{equation}
 where the  effective coupling, $g^{\rm Eff.}$, and the CP-violating (CPV) phase, $\delta_{\rm CP}$, are 
 \begin{equation}
  g^{Eff.} \equiv \sqrt{(y_{t}-g^{S})^{2}+g^{P~2}},\quad
  \delta_{CP} \equiv \sin^{-1}\left[ \frac{g^{P}}{\sqrt{(y_{t}-g^{S})^{2}+g^{P~2}}} \right].
 \end{equation}

\subsubsection{EFT Constraints}

\begin{table}
\caption{Bounds on EFT couplings}
 \begin{center}
\begin{tabular}{|c|c|c|}
\hline 
Operator & $g^{X}$ Bound & $\Lambda/\sqrt{\vert f_{\mathcal{O}}\vert}$ [GeV]  \tabularnewline\hline\hline 
\multirow{2}{*}{$\mathcal{O}_{t1}$} & $-0.72<g^{S}<0.21$   	& $>537$    \tabularnewline 
				    & $~~1.77<g^{S}<2.70$   	& $150 - 185$    \tabularnewline\hline
$\overline{\mathcal{O}}_{t1}$       & $-1.4<g^{P}<1.4$ 		& $>208$ \tabularnewline\hline
\end{tabular}
\label{eftBound.TB}
\end{center}
\end{table}

 Independent of deviations in the $h\rightarrow\gamma\gamma$ channel and with no assumption on the Higgs boson's total width,  
 ATLAS has measured the gluon-gluon fusion (ggF) scale factor to be~\cite{Chatrchyan:2013lba}
 \begin{equation}
  \kappa_g = 1.08^{+0.32}_{-0.14},\quad 
    \kappa_{g}^{2}\equiv \sigma(gg\rightarrow h)/\sigma^{\text{SM}}(gg\rightarrow h).
  \label{tthAnom.EQ}
  \end{equation}
Since ggF is dominated by a top quark loop, we can approximate an anomalous $g^{S}$ contribution to the observed rate by
\begin{equation}
 \sigma(gg\rightarrow h)= \kappa_{g}^{2}\times\sigma^{SM}(gg\rightarrow h)\approx \frac{(y_{t}-g^{S})^{2}}{y_{t}^{2}} \times\sigma^{SM}(gg\rightarrow h),
 \label{ggFApprox1.EQ}
\end{equation}
implying
\begin{equation}
 g^{S} \in [-0.72,0.21]\cup [1.77,2.70]~\text{at}~2\sigma.
 \label{gS2Sig.EQ}
\end{equation}
Similarly, we can relate Eq.~(\ref{tthAnom.EQ}) to $g^{P}$ by
 \begin{equation}
   \sigma(gg\rightarrow h)= \kappa_{g}^{2}\times\sigma^{SM}(gg\rightarrow h)\approx \frac{y_{t}^{2}+(g^{P})^{2}}{y_{t}^{2}} \times\sigma^{SM}(gg\rightarrow h),
   \label{ggFApprox2.EQ}
 \end{equation}
indicating
 \begin{equation}
 g^{P} \in [-1.41,1.41]~\text{at}~2\sigma.
 \label{gP2Sig.EQ}
\end{equation}
We next translate measurements of $\kappa_{g}$ into bounds on the cutoff scale of new physics involving operators 
$\mathcal{O}_{t1}$ and $\overline{\mathcal{O}}_{t1}$. 
The bounds on new physics scales $\Lambda/\sqrt{\vert f_{\mathcal{O}}\vert}$ are given in Table~\ref{eftBound.TB}. 
With the Naive Dimensional Analysis (NDA)~\cite{Manohar:1983md,Luty:1997fk} and $f_{\mathcal{O}}\sim \mathcal{O}(1)$, 
the new physics scale is pushed to about $\cal{O}$(1 TeV). 
Translating limits on $\kappa_{g}$ into bounds on $g^{L/R}$, and hence on $\overline{\mathcal{O}}_{\Phi q}^{(1)}$ and $\overline{\mathcal{O}}_{t2}$, 
is a nontrivial procedure due to the derivative coupling. 
Subsequently, such results are not presently available.


\subsection{Type I Two Higgs Doublet Model}
\label{2hdmI.SEC}
In the generic CP-conserving 2HDM, EWSB is facilitated by two $SU(2)_{L}$ doublets, $\Phi_{i}$, for $i\in\{1,2\}$, 
each with $U(1)_{Y}$ hypercharge $+1$ and a nonzero vacuum expectation value (vev) $v_{i}$. 
A $\mathcal{Z}_{2}$ symmetry is applied for $\Phi_1 \leftrightarrow \Phi_{2}$ to eliminate tree-level FCNC but may be softly broken at loop-level. 
After EWSB, there are five physical spin-0 states: $h,~H,~A,$ and $H^{\pm}$, 
which are respectively the two CP-even, single CP-odd, and $U(1)_{EM}$ charged Higgs bosons with masses $m_{h},m_{H}$, $m_{A}$, and $m_{H^\pm}$.
By convention, we fix the ordering of $h$ and $H$ by taking $$m_{h}<m_{H}.$$
Two angles, $\alpha$ and $\beta$, remain as free parameters.
$\alpha$ measures the mixing between the two CP-even Higgs fields to form the mass eigenstates ($h,\ H$)
and spans $\alpha\in[-\pi/2,\pi/2]$.
$\beta$ represents the relative size of $\langle\Phi_{i}\rangle$ and is defined by
\begin{equation}
 \tan\beta\equiv \langle\Phi_{2}\rangle / \langle\Phi_{1}\rangle= v_{2}/v_{1},  \quad \beta\in[0,\pi/2].
 \label{tB.EQ}
\end{equation}
Reviews of various 2HDMs and their phenomenologies can be found in Refs.~\cite{Gunion:1989we,Gunion:2002zf,Branco:2011iw}. 

\subsubsection{Type I 2HDM framework and parameters}
\label{2hdmITh.SEC}
In the 2HDM(I), much like in the SM, only one Higgs doublet is responsible for generating fermion masses and couples accordingly;
the second CP-even Higgs boson interacts with fermions through mixing.
The interaction Lagrangian relevant to this study is
\begin{eqnarray}
\mathcal{L}\ni
&-&\frac{gm_{u}}{2M_{W}}\overline{u}\left(h\frac{\cos\alpha}{\sin\beta}+H\frac{\sin\alpha}{\sin\beta}-i\gamma^{5}A\cot\beta\right)u\nonumber\\
&-&\frac{gm_{d}}{2M_{W}}\overline{d}\left(h\frac{\cos\alpha}{\sin\beta}+H\frac{\sin\alpha}{\sin\beta}+i\gamma^{5}A\cot\beta\right)d\nonumber\\
&+&gM_{W}W_{\mu}W^{\mu}\left[h\sin(\beta-\alpha)+H\cos(\beta-\alpha)\right]. 
\label{typeIlag.EQ}
\end{eqnarray} 
In Eq.~(\ref{typeIlag.EQ}), $u_{L(R)}$ is the LH (RH) up-type quark spinor, $d_{L(R)}$ is the down-type quark analogue, and
$g$ is the weak coupling constant in the SM.

\begin{table}
\caption{Neutral Scalar Boson Couplings in the 2HDM(I) Relative to the SM Higgs Couplings}
 \begin{center}
\begin{tabular}{|c|c|c|c|c|}
\hline 
Vertex & SM & 2HDM I & $\sin(\beta-\alpha)=1-\Delta_{V}$ \tabularnewline
\hline\hline 
$hu\overline{u}/d\overline{d}$ & $0$ or $1$ & $\frac{\cos\alpha}{\sin\beta}$  & $1-\Delta_{V}+\sqrt{2\Delta_{V}-\Delta_{V}^{2}}\cot\beta $ \tabularnewline\hline 
$hW^{+}W^{-}$    & $0$ or $1$ & $\sin(\beta-\alpha)$            & $1-\Delta_{V}$ \tabularnewline\hline 
$Hu\overline{u}/d\overline{d}$ & $0$ or $1$ & $\frac{\sin\alpha}{\sin\beta}$  & $(\Delta_{V}-1)\cot\beta+\sqrt{2\Delta_{V}-\Delta_{V}^{2}}$ \tabularnewline\hline 
$HW^{+}W^{-}$    & $0$ or $1$ & $\cos(\beta-\alpha)$            & $\sqrt{2\Delta_{V}-\Delta_{V}^{2}}$  \tabularnewline\hline 
$Au\overline{u}$ & - & $\cot\beta$                     & $\cot\beta$\tabularnewline\hline 
$Ad\overline{d}$ & - & $-\cot\beta$                    & $-\cot\beta$\tabularnewline\hline 
\end{tabular}
\label{2hdmICouple.TB}
\end{center}
\end{table}

Discovering a Higgs boson with SM-like couplings greatly impacts the 2HDM.
In particular, the measured couplings to weak bosons\cite{:2012gk,Chatrchyan:2013lba} imply either
\begin{eqnarray}
&& 
\sin(\beta-\alpha)\approx1\quad \text{for $h$ to be SM-like,}\\
\text{or}\ \ &&
\cos(\beta-\alpha)\approx1\quad \text{for $H$ to be SM-like.}
\label{sinBA.EQ}
\end{eqnarray}
Generally, we may parameterize how far $\sin(\beta-\alpha)$ is away from one and define $\Delta_{V}$ such that
\begin{equation}
\sin(\beta-\alpha)\equiv1-\Delta_{V},\quad0\leq\Delta_{V}\leq1.
\label{DVDef.EQ}
\end{equation}
We restrict the couplings to have the same sign as those of the SM \cite{Ellis:2013lra} and limit $\Delta_{V}$ up to one.
Eq.~(\ref{DVDef.EQ}) maps to the parameterization used by the SFitter Collaboration~\cite{Klute:2012pu} 
by taking $\Delta_{V}\rightarrow -\Delta_{V}$ and allowing $\Delta_{V}<0$.
After substituting $\alpha$ by $\Delta_{V}$ in Eq.~(\ref{typeIlag.EQ}), we have

{\small
\begin{eqnarray}
\mathcal{L}\ni
&-&\frac{gm_{u}}{2M_{W}}\overline{u}\left[h\left(1-\Delta_{V}+\sqrt{2\Delta_{V}-\Delta_{V}^{2}}\cot\beta\right)+H\left((\Delta_{V}-1)\cot\beta+\sqrt{2\Delta_{V}-\Delta_{V}^{2}}\right)\right]u\nonumber\\
&-&\frac{gm_{d}}{2M_{W}}\overline{d}\left[h\left(1-\Delta_{V}+\sqrt{2\Delta_{V}-\Delta_{V}^{2}}\cot\beta\right)+H\left((\Delta_{V}-1)\cot\beta+\sqrt{2\Delta_{V}-\Delta_{V}^{2}}\right)\right]d\nonumber\\
&+&\frac{gm_{u}}{2M_{W}}\overline{u}\left[i\gamma^{5}A\cot\beta\right]u - \frac{gm_{d}}{2M_{W}}\overline{d}\left[i\gamma^{5}A\cot\beta\right]d\nonumber\\
&+&gM_{W}W_{\mu}W^{\mu}\left[h(1-\Delta_{V})+H\sqrt{2\Delta_{V}-\Delta_{V}^{2}}\right].
\label{typeIlagDV.EQ}
\end{eqnarray} }
Table \ref{2hdmICouple.TB} summarizes the bosonic and fermionic couplings to the neutral scalar in the 2HDM(I) relative to those in the SM, i.e., the 2HDM(I) coupling coefficient divided by the SM coupling coefficient.
In the small (large) $\Delta_{V}$ limit, $h~(H)$ becomes SM-like and $H~(h)$ becomes non-SM-like.
At $\Delta_{V}=0~(\Delta_{V}=1)$, $H~(h)$ decouples from the gauge bosons. 
The relevant tree-level couplings to $A$ are independent of $\Delta_{V}$ as they are initially independent of $\alpha$. 
In the large $\tan\beta$ limit, $A$ decouples from the theory.
For all parameter scenarios considered, we identify the SM-like Higgs as the one with stronger couplings to $WW,\ ZZ$, and having a mass of $125.5$ GeV.

\subsubsection{Type I 2HDM Constraints}
\label{2hdmIConst.SEC}
Since the Higgs boson's discovery, many reports have appeared investigating the 2HDMs' compatibility with 
data~\cite{Ellis:2013lra,Klute:2012pu,Baak:2011ze,Gorczyca:2011he,Ferreira:2012my,Altmannshofer:2012ar,Azatov:2012qz,Chen:2013kt,Chiang:2013ixa,Chang:2013aya,Barroso:2013zxa,Chen:2013rba,Craig:2013hca,Barger:2013mga,Mahmoudi:2012ej}.
We list here constraints relevant to the 2HDM(I) and note when a result is applicable to other types.
The following bounds assume one SM-like Higgs boson at approximately $126$ GeV.

\begin{enumerate}[(i)]
 \item{\it $\cos(\beta-\alpha)-\tan\beta$ Parameter Space}:
    A global fit of available LHC data, in particular from $h \to \gamma\gamma, \ VV,\ b\bar b,\ \tau^{+}\tau^{-}$,  
    has set stringent bounds \cite{Craig:2013hca}. Representative values at $95\%$CL are
    \begin{equation}
     \cos(\beta-\alpha)< 0.3~(0.40)~[0.42] \quad\text{for}\quad \tan\beta=2.4~(10)~[100].
     \label{2hdmICostB.EQ}
    \end{equation}
    Similar conclusions have been reached by Refs.~\cite{Azatov:2012qz,Chiang:2013ixa,Barger:2013mga,Barroso:2013zxa,Chen:2013rba}.    
   
  \item{\it $m_{H^{\pm}}-\tan\beta$ Parameter Space}:
  For all 2HDMs, flavor observables exclude at 95\%~CL~\cite{Barberio:2008fa,Mahmoudi:2009zx}
  \begin{equation}
   \tan\beta < 1 \quad\text{for}\quad  m_{H^{\pm}}<500~\text{GeV}.
  \end{equation}
  Values of $\tan\beta<1$ are allowed given a sufficiently heavy $H^{\pm}$~\cite{Barberio:2008fa,Mahmoudi:2009zx,Mahmoudi:2012ej,Chen:2013kt}.
  Due to the particular $\tan\beta$ dependence, 
  no absolute lower bound on $m_{H^{\pm}}$ from flavor constraints exists in the 2HDM(I)~\cite{Mahmoudi:2009zx}.
  An observation of excess $B\rightarrow D^{*}\tau\nu$ decays~\cite{Lees:2013uzd} has yet to be confirmed and is not considered.
  
  \item {\it Additional Higgs Masses}: 
  For both 2HDM(I) and (II), additional CP-even scalars below LEP bounds~\cite{Barate:2003sz,Abbiendi:2004gn,Abdallah:2004wy} are allowed 
  given sufficiently decoupled $H^{\pm}$ and $A$~\cite{Ferreira:2012my}.
  A second CP-even Higgs is incompatible with LHC data for mass
  \begin{equation}
  180~\text{GeV}< m_{H} < 350~\text{GeV},
  \end{equation}
  but allowed outside this range~\cite{Altmannshofer:2012ar}.
  Direct searches for $H^{\pm}$ and $A$ exclude ~\cite{Abbiendi:2004gn,Aubert:2009cp,Abdallah:2004wy,Abbiendi:2013hk}
   \begin{equation}
   m_{H^\pm},~m_{A} \lesssim 80~\text{GeV}.
   \end{equation}

\end{enumerate}

 Additional considerations include the compatibility of a  
SM-like Higgs boson with EW precision data in general 2HDMs \cite{Baak:2011ze},
the perturbative unitarity limits on the heavy Higgs masses in a general, CP-conserving 2HDM 
\cite{Kanemura:1993hm,Akeroyd:2000wc,Chang:2013aya}, 
and perturbative unitarity limits on $\tan\beta$ in an exact $\mathcal{Z}_{2}$-symmetric, CP-conserving 2HDM~\cite{Gorczyca:2011he,Barroso:2013zxa}.
Since FCNC do exist in nature and the SM, it is unnecessary to impose the severe constraints on $\tan\beta$ associated with an exact  $\mathcal{Z}_{2}$ symmetry.


\subsection{Type II Two Higgs Doublet Model}
\label{2hdmII.SEC}

\subsubsection{Type II 2HDM framework and parameters}
\label{2hdmIITh.SEC}
In the 2HDM(II), one Higgs doublet is assigned a hypercharge $+1$, giving masses to fermions with weak isospin $T^{3}_{L}=+\frac{1}{2}$, 
and the second is assigned a hypercharge $-1$, giving masses to $T^{3}_{L}=-\frac{1}{2}$ fermions.
The doublets are denoted respectively by $\Phi_{u}$ and $\Phi_{d}$, and $\beta$ is written as
\begin{equation}
\tan\beta\equiv\langle \Phi_{u}\rangle / \langle \Phi_{d}\rangle  = v_{u}/v_{d}.
\end{equation}
After EWSB, the CP-conserving interaction Lagrangian relevant to Eq.~(\ref{tWbH.EQ}) is similar to Eq.~(\ref{typeIlag.EQ}), 
with the only difference being the down-type quark Yukawa couplings:
\begin{eqnarray}
\mathcal{L}\ni
&-&\frac{gm_{d}}{2M_{W}}\overline{d}\left(-h\frac{\sin\alpha}{\cos\beta}+H\frac{\cos\alpha}{\cos\beta}-i\gamma^{5}A\tan\beta\right)d.
\label{typeIILag.EQ}
\end{eqnarray}
The notation used in Eq.~(\ref{typeIILag.EQ}) is the same as the 2HDM(I) Lagrangian Eq.~(\ref{typeIlag.EQ}).
Using Eq.~(\ref{DVDef.EQ}), and similar to Eq.~(\ref{typeIlagDV.EQ}), the preceding line becomes 
{\small
\begin{eqnarray}
\mathcal{L}\ni 
&-&\frac{gm_{d}}{2M_{W}}\overline{d}\left[h\left(1-\Delta_{V}-\sqrt{2\Delta_{V}-\Delta_{V}^{2}}\tan\beta\right)+H\left((1-\Delta_{V})\tan\beta+\sqrt{2\Delta_{V}-\Delta_{V}^{2}}\right)\right]d\nonumber\\
&+& i \frac{gm_{d}}{2M_{W}}\overline{d}\gamma^{5}d \ A\tan\beta .
\label{lagDV.EQ}
\end{eqnarray} }
Table \ref{2HDMII.TB} summarizes the bosonic and fermionic couplings to the neutral scalars in the 2HDM(II) relative to those in the SM.
Like the 2HDM(I), in the small (large) $\Delta_{V}$ limit, $h~(H)$ becomes SM-like and $H~(h)$ becomes non-SM-like.
At $\Delta_{V}=0~(\Delta_{V}=1)$, $H~(h)$ decouples from the gauge bosons. 
In this same limit, the $h~(H)$ Yukawa couplings become independent of $\tan\beta$. 
Unlike the 2HDM(I), $A$ only  decouples from the theory if taken to be infinitely heavy.

An important feature for the Higgs couplings to fermions is that 
the down-type quark couplings are enhanced at higher values of $\tan\beta$, while 
the up-type quark couplings are suppressed. For the charged Higgs however, there is an interplay between the two and the particular value $\tan\beta =\sqrt{m_{t}^{\overline{\text{MS}}}(m_{t})/m_{b}^{\overline{\text{MS}}}(m_{t})}\ \approx 7.6$ minimizes the decay $t\rightarrow H^{+}b$.
Though no such minima occur in the 2HDM(I), sensitivity to $\tan\beta=7.6$ will be investigated in both 2HDM scenarios.

\begin{table}
\caption{Neutral Scalar Boson Couplings in the 2HDM(II) Relative to the SM Higgs Couplings}
 \begin{center}
\begin{tabular}{|c|c|c|c|c|}
\hline 
Vertex & SM & 2HDM II & $\sin(\beta-\alpha)=1-\Delta_{V}$ \tabularnewline
\hline\hline 
$hu\overline{u}$ & $0$ or $1$ & $\frac{\cos\alpha}{\sin\beta}$  & $1-\Delta_{V}+\sqrt{2\Delta_{V}-\Delta_{V}^{2}}\cot\beta $ \tabularnewline\hline 
$hd\overline{d}$ & $0$ or $1$ & $-\frac{\sin\alpha}{\cos\beta}$ & $1-\Delta_{V}-\sqrt{2\Delta_{V}-\Delta_{V}^{2}}\tan\beta $ \tabularnewline\hline 
$hW^{+}W^{-}$    & $0$ or $1$ & $\sin(\beta-\alpha)$            & $1-\Delta_{V}$ \tabularnewline\hline 
$Hu\overline{u}$ & $0$ or $1$ & $\frac{\sin\alpha}{\sin\beta}$  & $(\Delta_{V}-1)\cot\beta+\sqrt{2\Delta_{V}-\Delta_{V}^{2}}$ \tabularnewline\hline 
$Hd\overline{d}$ & $0$ or $1$ & $\frac{\cos\alpha}{\cos\beta}$  & $(1-\Delta_{V})\tan\beta+\sqrt{2\Delta_{V}-\Delta_{V}^{2}}$ \tabularnewline\hline 
$HW^{+}W^{-}$    & $0$ or $1$ & $\cos(\beta-\alpha)$            & $\sqrt{2\Delta_{V}-\Delta_{V}^{2}}$  \tabularnewline\hline 
$Au\overline{u}$ & $-$ & $\cot\beta$                     & $\cot\beta$\tabularnewline\hline 
$Ad\overline{d}$ & $-$ & $\tan\beta$                     & $\tan\beta$\tabularnewline\hline 
  \end{tabular}
\label{2HDMII.TB}
\end{center}
\end{table}

\subsubsection{Type II 2HDM Constraints}
\label{2hdmIIConst.SEC}
Constraints relevant to the 2HDM(II) are listed here.
See Sec.~\!\!\!\!\!\!\!\!\!\!\ref{2hdmI.SEC} for generic 2HDM bounds.

\begin{enumerate}[(i)]

\item{\it $\cos(\beta-\alpha)-\tan\beta$ Parameter Space}:
    A global fit of available LHC data, in particular from $h \to \gamma\gamma, \ VV,\ b\bar b,\ \tau^{+}\tau^{-}$,  
    has set stringent bounds \cite{Craig:2013hca}. Representative values at $95\%$~CL are
    \begin{equation}
     \cos(\beta-\alpha)< 0.06~(0.01) \quad\text{for}\quad \tan\beta=2.4~(10).
     \label{2hdmIICostB.EQ}
    \end{equation}
    Similar conclusions have been reached by Refs.~\cite{Azatov:2012qz,Chiang:2013ixa,Barger:2013mga,Barroso:2013zxa,Chen:2013rba}.

  \item {\it $m_{H^{\pm}}-\tan\beta$ Parameter Space}: 
  Flavor observables, and in particular $\BR(B\rightarrow X_{s}\gamma)$, exclude at 95\%~CL~\cite{Misiak:2006zs,Barberio:2008fa,Lees:2012ym}
  \begin{equation}
    m_{H^{\pm}}<327~\text{GeV}\quad\text{for all}~\tan\beta
  \end{equation}
   From $\BR(B\rightarrow\tau\nu)$ measurements, the {\bf UT}{\it fit} Collaboration~\cite{Bona:2009cj} has determined the absolute bound
 \begin{equation}
  \tan\beta < 7.4 \frac{m_{H^\pm}}{100~\text{GeV}}.
 \end{equation}

 \end{enumerate}

\section{BRANCHING RATIOS}
\label{br.SEC}
The discovery of a SM-like Higgs boson at 126 GeV\cite{:2012gk,Chatrchyan:2013lba} implies that
\begin{equation}
 t\rightarrow W^{+*}bh,\quad
  W^{+*}\rightarrow f_{1}\bar f_{2}
 \label{tWbhWmv.EQ}
\end{equation}
is kinematically allowed and proceeds through the diagrams given in Fig.~\ref{feynman.FIG}. 
Following Ref.~\cite{Mahlon:1994us}, we define the $t\rightarrow W^{*}bh$ partial width as
\begin{equation}
 \Gamma(t\rightarrow Wbh)= \frac{\Gamma(t\rightarrow \mu^{+}\nu_{\mu}bh)}{\BR(W\rightarrow\mu\nu_{\mu})},
 \label{GWbh.EQ}
\end{equation}
and the $t\rightarrow W^{*}bh$ branching ratio by
\begin{equation}
 \BR(t\rightarrow Wbh)= \frac{\Gamma(t\rightarrow Wbh)}{\Gamma_{\rm Tot.}},\quad \Gamma_{\rm Tot.}\equiv \Gamma(t\rightarrow Wb).
 \label{brWbh.EQ}
\end{equation}

With CalcHEP 3.4.2~\cite{Pukhov:1999gg}, we find excellent agreement with Ref.~\cite{Mahlon:1994us}. 
With updated parameters~\cite{:2012gk,Chatrchyan:2013lba,Beringer:1900zz}:
{\small
\begin{eqnarray}
 m_{t}^{\overline{\text{MS}}}(m_{t})=173.5~\text{GeV},\quad
 m_{b}^{\overline{\text{MS}}}(m_{t})=3.01~\text{GeV},\quad
 m_{h}&=&125.5~\text{GeV},\quad
 m_{\mu}=0~\text{GeV},
 \nonumber\\
 M_{W}=80.385~\text{GeV}, \quad
 M_{Z}=91.1876~\text{GeV},  \quad
 G_{F}&=&1.1663787\times10^{-5}~\text{GeV}^{-1},
 \nonumber\\  
 \Gamma_{W}=2.085~\text{GeV},\quad
 \BR(W\rightarrow\mu\nu)&=&0.1057,
 \label{input.EQ}
\end{eqnarray} }
we calculate $\Gamma_{\rm Tot.}$ at leading order to be
\begin{equation}
 \Gamma_{\rm Tot.} = 1.509~\text{GeV},
 \label{totWidth.EQ}
\end{equation}
and find that the SM predicts
\begin{equation}
 \BR^{\rm SM}(t\rightarrow Wbh) = 1.80\times10^{-9}.
 \label{smBR.EQ}
 \end{equation} 
The smallness of this branching fraction 
falls from several features, including phase space suppression associated with the three-body final state, kinematic suppression due to the off-shell $W$ boson, 
and an accidental cancellation between the leading $t\overline{t}h$ and subleasing $WWh$ diagrams.
Nevertheless, this decay rate is $\mathcal{O}(10^{5})$ larger than the well-studied~\cite{Eilam:1990zc,Mele:1998ag} two-body $t\rightarrow ch$ transition.
This is due to the GIM suppression for the FCNC. 
 
In the remainder of this section, we investigate how the branching fraction can change in the context of EFT, 2HDM(I), and 2HDM(II).
  

\subsection{EFT BR$(t\rightarrow Wbh)$}
\label{EffBR.SEC}

 \begin{table}
\caption{$\BR(t\rightarrow Wbh)$ for Benchmark Values of Anomalous $t\overline{t}h$ Couplings}
 \begin{center}
\begin{tabular}{|c|c|c|}
\hline 
 \multicolumn{2}{|c|}{$g^{X}$} & $\BR(t\rightarrow Wbh)$  \tabularnewline\hline\hline 
 $g^{S}$& $~0.5$ & $1.075\times 10^{-9}$ \tabularnewline\hline
  & $-0.5$ & $3.078\times 10^{-9}$  \tabularnewline\hline
 $g^{P}$& $~0.5$ & $1.929\times 10^{-9}$  \tabularnewline\hline
  & $-0.5$ & $1.928\times 10^{-9}$  \tabularnewline\hline
 $g^{L}$& $~0.5$ & $1.812\times 10^{-9}$  \tabularnewline\hline 
  & $-0.5$ & $1.812\times 10^{-9}$  \tabularnewline\hline
 $g^{R}$& $~0.5$ & $1.927\times 10^{-9}$  \tabularnewline\hline
  & $-0.5$ & $1.928\times 10^{-9}$  \tabularnewline\hline\hline
\end{tabular}
\label{eftBench.TB}
\end{center}
\end{table}

\begin{figure}[tb]
\centering
\subfigure[]{
	\includegraphics[width=0.46\textwidth]{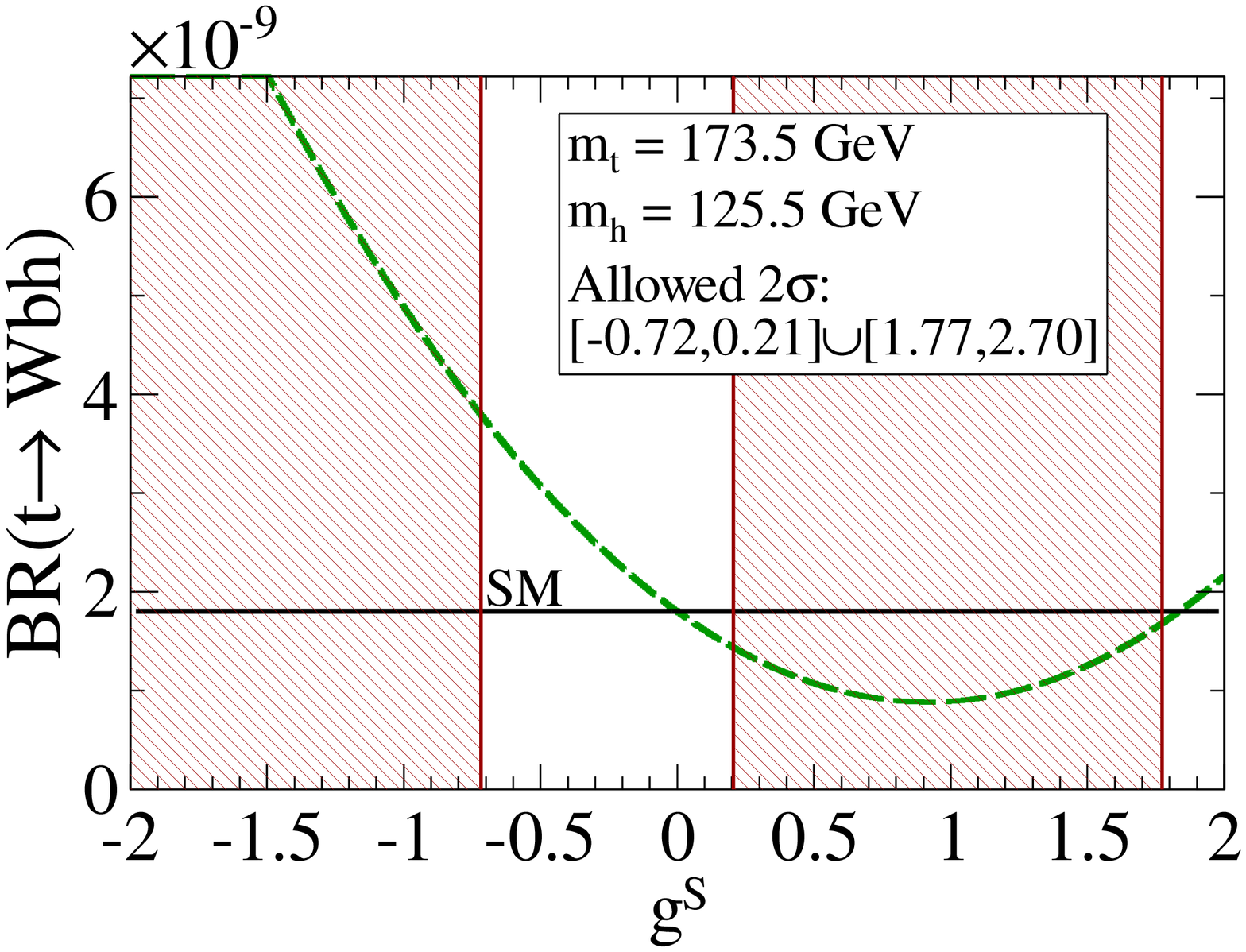}	
	\label{gs.FIG}
	}
\subfigure[]{
	\includegraphics[width=0.46\textwidth]{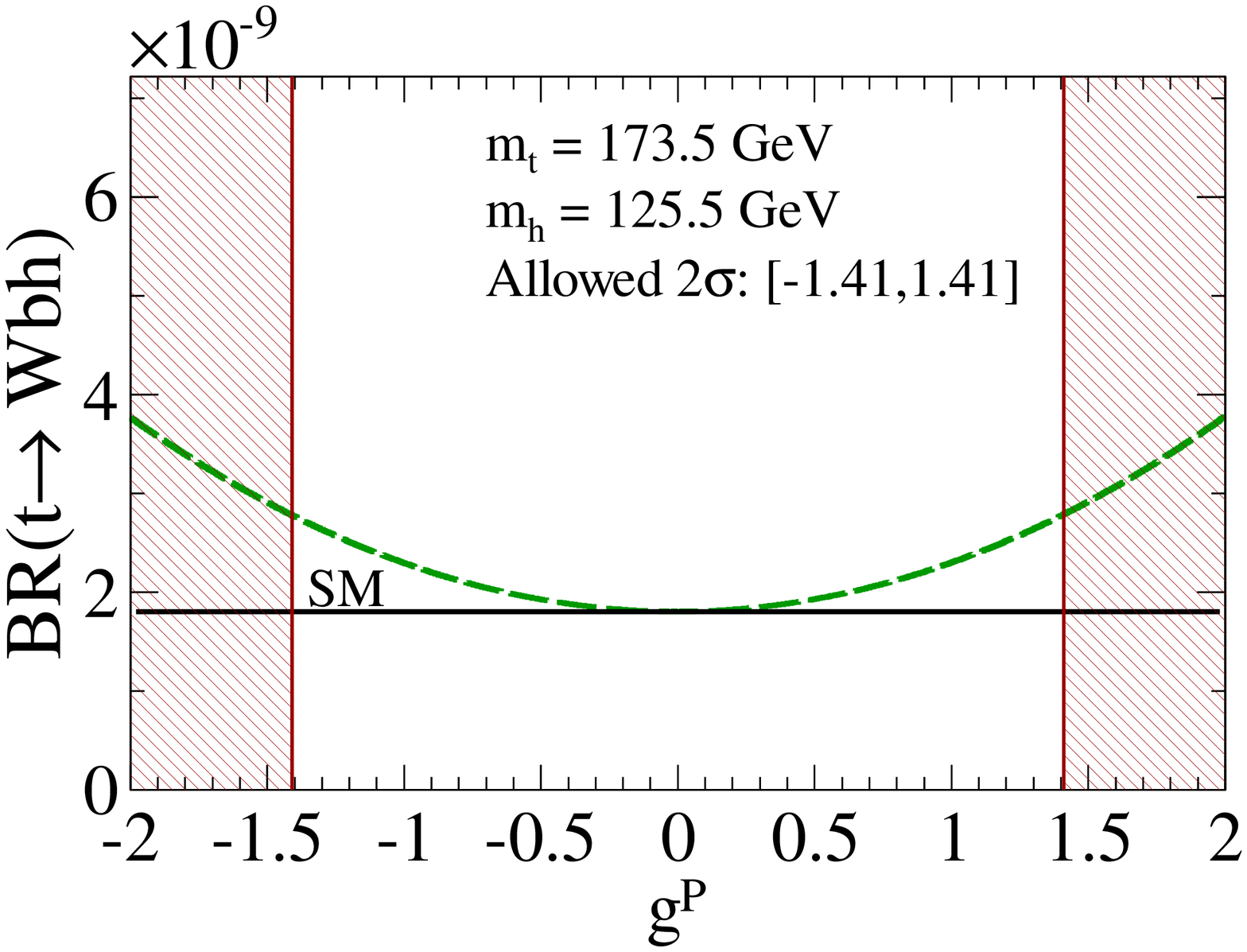}	
	\label{gp.FIG}
	}
\\
\subfigure[]{
	\includegraphics[width=0.46\textwidth]{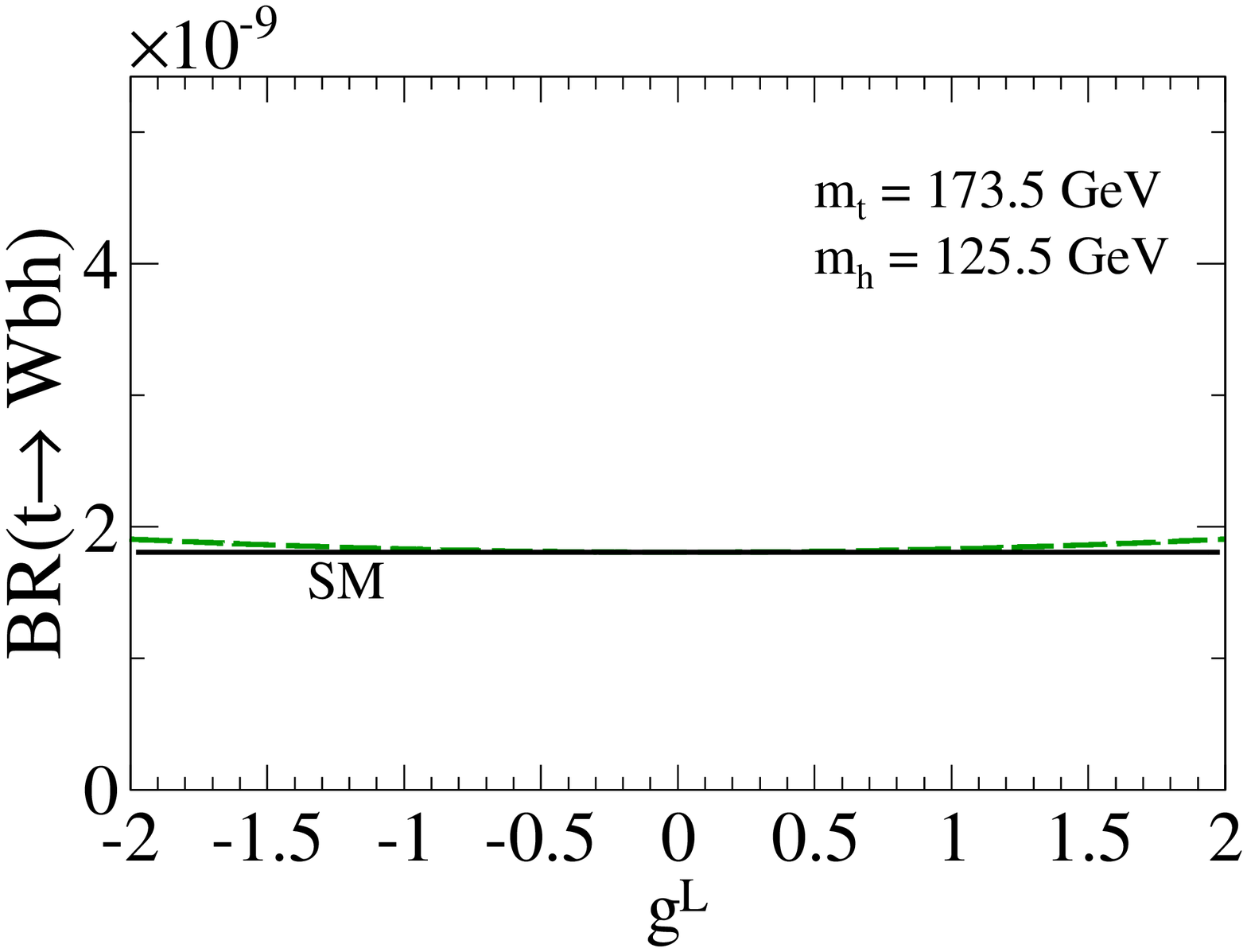}
	\label{gl.FIG}
	}
\subfigure[]{
	\includegraphics[width=0.46\textwidth]{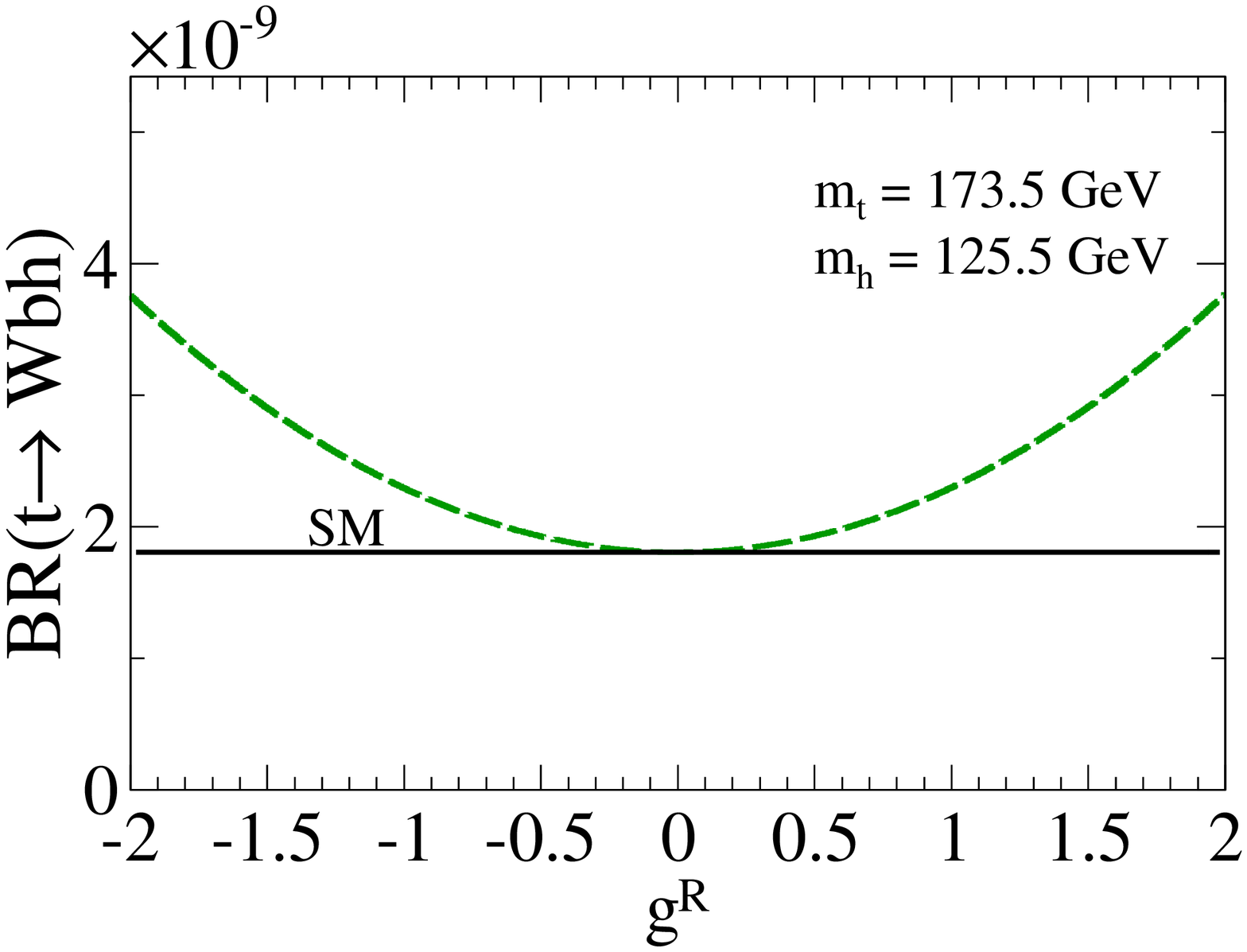}	
	\label{gr.FIG}
}
\caption{
$\BR(t\rightarrow Wbh)$ as a function of
(a) $g^{S}$, (b) $g^{P}$, (c) $g^{L}$, (d) $g^{R}$.
The solid line denotes the SM prediction, Eq.~(\ref{smBR.EQ}).
The shaded region is excluded at 95\% C.L.
}
\label{effBR.FIG}
\end{figure}

We present first the behavior of  $\BR(t\rightarrow Wbh)$ as a function of anomalous $t\overline{t}h$ couplings.
For one non-zero anomalous coupling from Eq.~(\ref{eftLag.EQ}) at a time, we calculate the branching fraction over the domain $g^{X}\in[-2,+2]$ 
and set all other anomalous couplings to zero.
Bounds on $g^{S}$ and $g^{P}$, Eqs.~(\ref{gS2Sig.EQ}) and (\ref{gP2Sig.EQ}) respectively, are applied. 
The results are shown in Fig.~\ref{effBR.FIG}.
To investigate the sensitivity of operators that select out different kinematic features, 
we include the redundant operators listed in Eq.~(\ref{GIOp2.EQ}), which give rise to anomalous $g^{L}$ and $g^{R}$. 
In all plots, the SM prediction  as in Eq.~(\ref{smBR.EQ}) is shown as a (black) solid line labeled by ``SM''.
Table~\ref{eftBench.TB} lists values of the branching fraction for various benchmark values of $g^{X}$.

In Fig.~\ref{gs.FIG}, $\BR(t\rightarrow Wbh)$ as a function of the anomalous scalar coupling $g^{S}$ is shown. 
From the Lagrangian in Eq.~(\ref{cpvLag.EQ}), it is clear that $(y_{t}-g^{S})$ acts as an  effective Yukawa coupling.
For $g^{S}<0$, the anomalous coupling enhances the already dominant top-Higgsstrahlung diagram.
For $g^{S}>0$, an accidental cancellation among the anomalous scalar, Yukawa, and gauge terms results in a minimum at $g^{S}\approx0.92$.
When $g^{S}\gtrsim0.92$, the quadratic term takes over and causes the branching fraction to grow.
An observed transition rate smaller than the SM prediction thus implies that $g^{S}>0$.
Indirect measurements of the $t\overline{t}h$ coupling, as seen in Fig.~\ref{gs.FIG}, indicate that 
\begin{equation}
\BR(t\rightarrow Wbh) = (0.8\sim2.1) \times \BR^{\rm SM}(t\rightarrow Wbh).
\label{eftBRLimitI.EQ}
\end{equation}

Figure~\ref{gp.FIG} shows the influence of an anomalous pseudoscalar coupling, $g^{P},$ on $\BR(t\rightarrow Wbh)$.
From the Lagrangian in Eq.~(\ref{cpvLag.EQ}), similar to the discussions in the previous session, the $t\rightarrow h$ transition is symmetric with respect to $g^{P}$ due to the dominance of the quadratic term.
Both couplings contribute greatest when the intermediate, off-shell top quark propagates in its RH helicity state, which gives an $m_{t}$ enhancement over other diagrams.
The CPV associated with $\delta_{\rm CP}$ is unobservable here because the asymmetry is proportional to interference terms, which are small. 

The  linear dependence on $g^{S}$ in interference terms from the previous case and the strict quadratic dependence on $g^{P}$ here implies that that branching fraction is less sensitive to small values of $g^{P}$ than it is to small values of $g^{S}$.
The rate therefore grows more slowly as a function of  $g^{P}$ than $g^{S}$.
As seen in Figure~\ref{gp.FIG}, the bounds on $g^{P}$ allow
\begin{equation}
\BR(t\rightarrow Wbh) = (1\sim 1.5) \times \BR^{\rm SM}(t\rightarrow Wbh).
\end{equation}

In Fig.~\ref{gl.FIG}, we see the branching fraction as a function of an anomalous LH vector current with coupling $g^{L}$.
Over the domain investigated, the contribution is rather small.
We turn to kinematics to elucidate this behavior.
First, the anomalous contribution is proportional to $k_\mu/v$, where $k_{\mu}$ is the momentum of the Higgs. 
Since the energy budget for this process is fixed at $m_{t}$, and since we require a final state Higgs $(E_{h}\gtrsim m_{h})$, 
$k_\mu/v\sim E_{h}/v$ ranges between $0.5\sim0.6$, leading to kinematic suppression of anomalous contributions.
Second, note that a fermion participating in two sequential LH chiral interactions necessarily propagates in its LH helicity state.
Hence, the anomalous contribution is proportional to the internal, off-shell top quark momentum and leads to helicity suppression of anomalous contributions. 
We consequently expect and observe very small growth in the branching fraction over the range of $g^{L}$.

Figure~\ref{gr.FIG} displays the results for $\BR(t\rightarrow Wbh)$ as a function of anomalous RH vector current with coupling $g^{R}$.
Unlike the LH case, the anomalous contribution has a large effect over the domain considered, comparable to $g^{S}$ and $g^{P}$.
As in the previous case, there is kinematic suppression; however, there is no longer helicity suppression.
A massive fermion participating in a RH chiral interaction followed by a LH chiral interaction propagates in its RH helicity state.
Hence, as in the $g^{P}$ case, the anomalous contribution is proportional to $m_{t}$. 
Comparatively, there is a faster rise in the transition rate as a function of $g^{R}$ than $g^{L}$.

 \begin{table}
\caption{$\BR(t\rightarrow WbH)$ for Benchmark Values of Higgses in the 2HDM(I)}
 \begin{center}
\begin{tabular}{|c|c|c|c|c|}
\hline 
 $H~(125.5~{\rm GeV})$ &  $\Delta_{V}$ & $\tan\beta$ & $\BR^{2HDM(I)}(t\rightarrow WbH)$  \tabularnewline\hline\hline 
 $h$ &  $0.05$ & $3$   & $1.840 \times 10^{-9}$     \tabularnewline\hline
 $h$ &         & $7.6$ & $1.714 \times 10^{-9}$     \tabularnewline\hline
 $H$ &  $0.7$  & $3$   & $1.460 \times 10^{-9}$     \tabularnewline\hline 
 $H$ &         & $7.6$ & $1.567 \times 10^{-9}$     \tabularnewline\hline
 $h$ &  $0.7$  & $3$   & $4.643 \times 10^{-10}$    \tabularnewline\hline
 $h$ &         & $7.6$ & $2.573 \times 10^{-10}$    \tabularnewline\hline
 $A~(100~{\rm GeV})$ &      & $3$   & $1.814 \times 10^{-9}$    \tabularnewline\hline
 $A~(100~{\rm GeV})$ &      & $7.6$ & $2.829 \times 10^{-10}$  \tabularnewline\hline\hline\end{tabular}
\label{2hdmIBench.TB}
\end{center}
\end{table}

\begin{figure}[tb]
\centering
\subfigure[]{
	\includegraphics[width=0.47\textwidth]{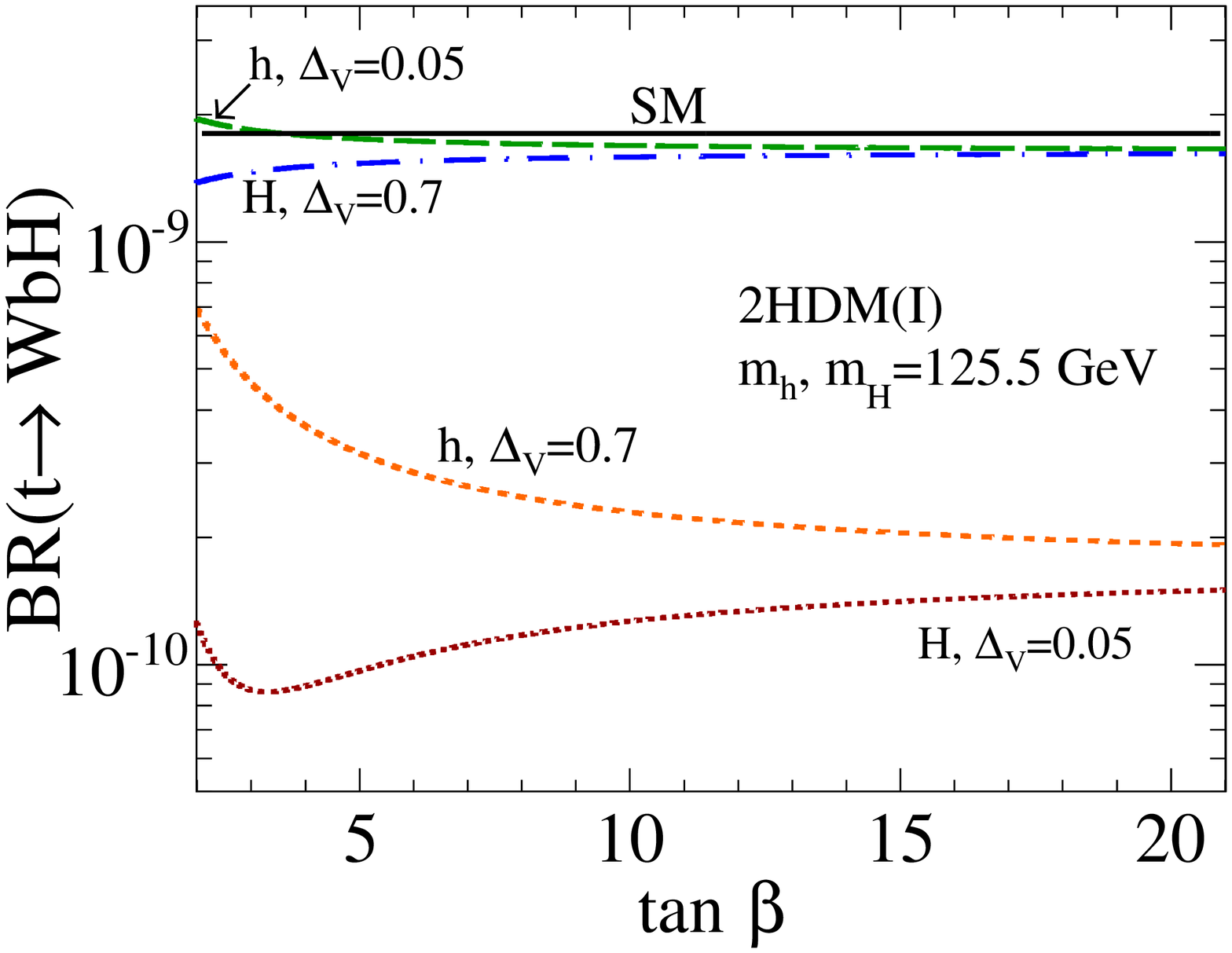}	
	\label{2hdmI_brVstB.FIG}
}
\subfigure[]{
	\includegraphics[width=0.47\textwidth]{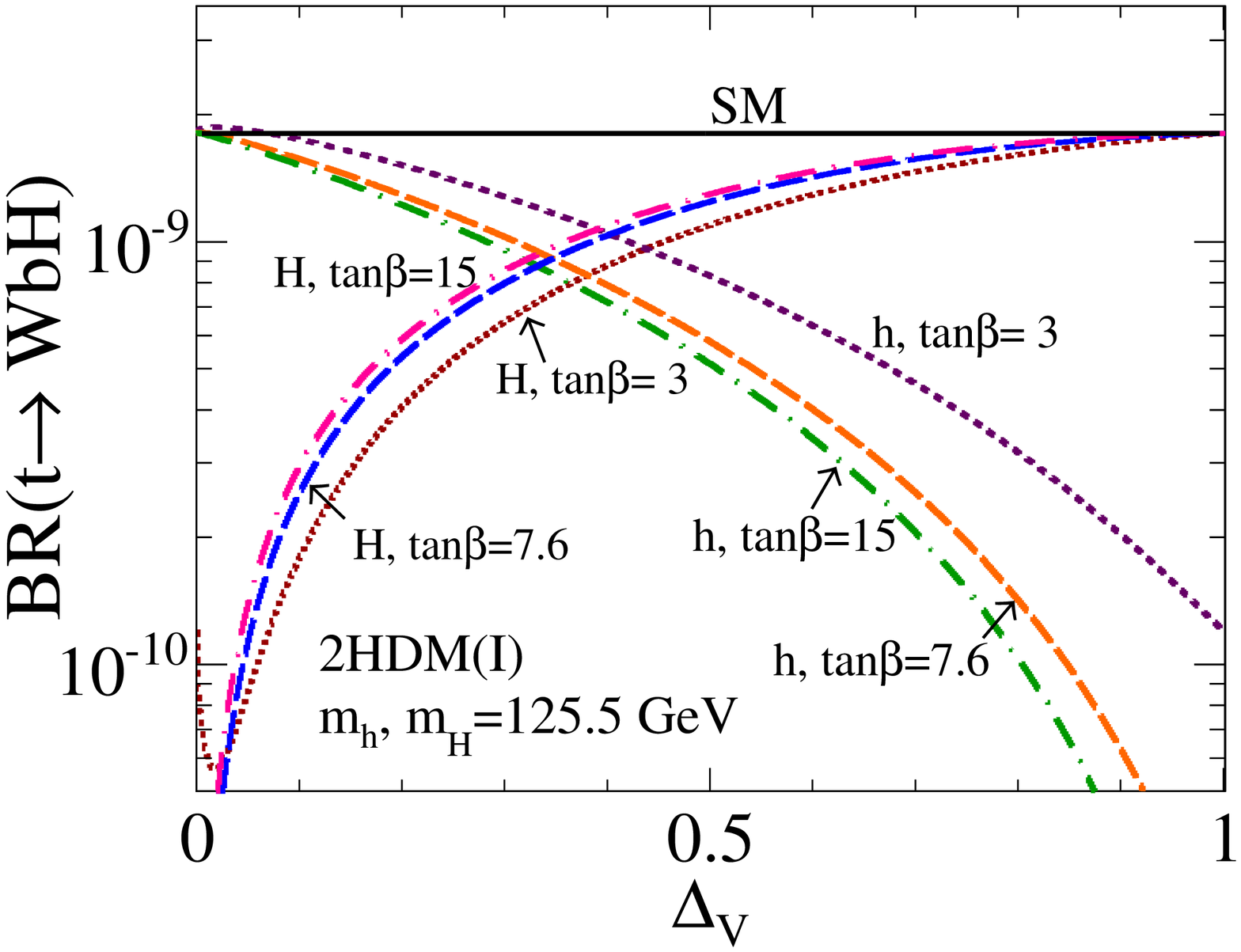}	
	\label{2hdmI_brVsDV.FIG}
}
\caption{
The 2HDM(I) $\BR(t\rightarrow WbH)$ as a function of 
(a) $\tan\beta$ for SM-like $h$ (long dash), $H$ (dash-dot), and non-SM-like $h$ (short dash), $H$ (dot);
(b) $\Delta_{V}$ for $h$ at $\tan\beta = 3~7.6,~15$ (short dash, long dash, dash-dot), and for $H$ (dot, long dash, dash-dot).
The solid line denotes the SM prediction, Eq.~(\ref{smBR.EQ}).
}
\label{2hdmI_HvsInput.FIG}
\end{figure}

\subsection{Type I 2HDM BR$(t\rightarrow WbH)$}
\label{2hdmIBR.SEC}
The behavior of $\BR(t\rightarrow WbH)$, where $H$ represents $h,~H,$ or $A$ in the 2HDM(I), is presented in this section.
To explore sensitivity to the anomalous $WWH$ coupling, $\Delta_{V}$, we consider
\begin{equation}
 \tan\beta=3,~7.6,~15\quad\text{for}~\Delta_{V}\in[0,1].
\end{equation}
For these values of $\tan\beta$, 
the largest deviation in the $WWH$ coupling allowed by present data corresponds to a light SM-like Higgs with $\cos(\beta-\alpha)=0.3$, i.e.,
\begin{equation}
  \Delta_{V}=0.05~(0.7)\quad\text{for}\quad h~(H)\approx h^{SM}.
\end{equation}
To determine the mass sensitivity, we focus on the mass windows 
\begin{eqnarray}
  m_{h}\in[95~\text{GeV},126~\text{GeV}]\label{mhMass.EQ},\ \
  m_{H}\in [126~\text{GeV},155~\text{GeV}]\label{mHMass.EQ},\ \
  m_{A}\in [95~\text{GeV},155~\text{GeV}]\label{mAMass.EQ}.
\end{eqnarray}
Below 95 GeV, the SM $Z$ boson background becomes relevant, making observation of the transition very difficult;
above $155$ GeV the kinematic suppression of $t\rightarrow H/A$ becomes too great for practical purposes.
However, it is straightforward to extrapolate these results in the event of a neutral scalar's discovery in these peripheral ranges.

Table~\ref{2hdmIBench.TB} lists values of $\BR(t\rightarrow WbH)$ for several Higgses and benchmark parameter values.

\subsubsection{BR$(t\rightarrow Wbh, H)$ vs $\tan\beta$, $\Delta_{V}$}

\begin{figure}[tb]
\centering
\subfigure[]{
	\includegraphics[width=0.47\textwidth]{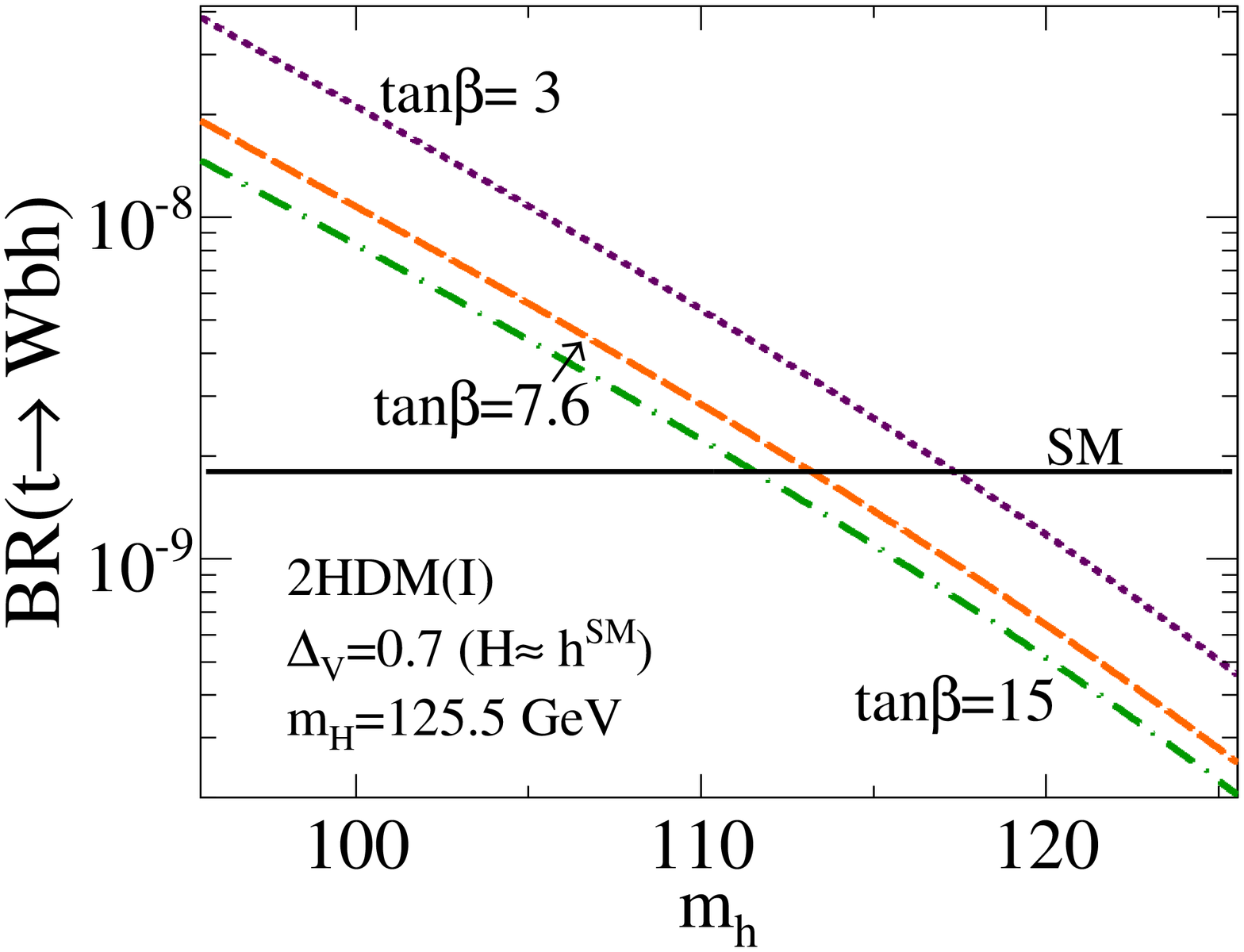}
	 \label{2hdmI_brVsMH1.FIG}
}
\subfigure[]{
	\includegraphics[width=0.47\textwidth]{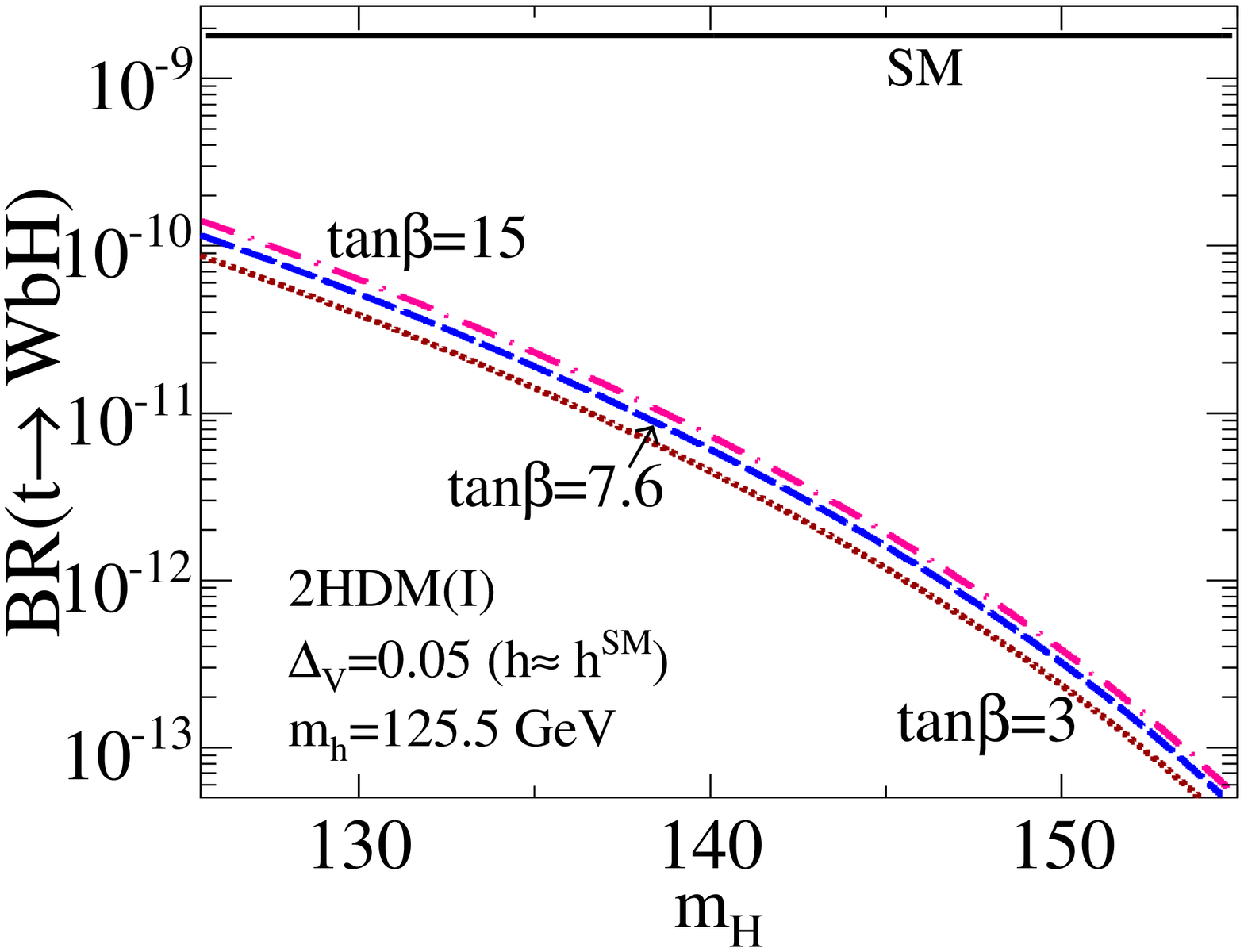}	
	\label{2hdmI_brVsMH2.FIG}
}
\vspace{.15in}\\
\caption{
The 2HDM(I) $\BR(t\rightarrow WbH)$ as a function of mass for a non-SM-like 
(a) $h$ and (b) $H$ assuming $\tan\beta=3,~7.6,~15$ (short dash, long dash, dash-dot).
The solid line denotes the SM prediction, Eq.~(\ref{smBR.EQ}).
}
\label{2hdmI_HvsMH.FIG}
\end{figure}

\begin{figure}[tb]
\centering
\subfigure[]{
	\includegraphics[width=0.47\textwidth]{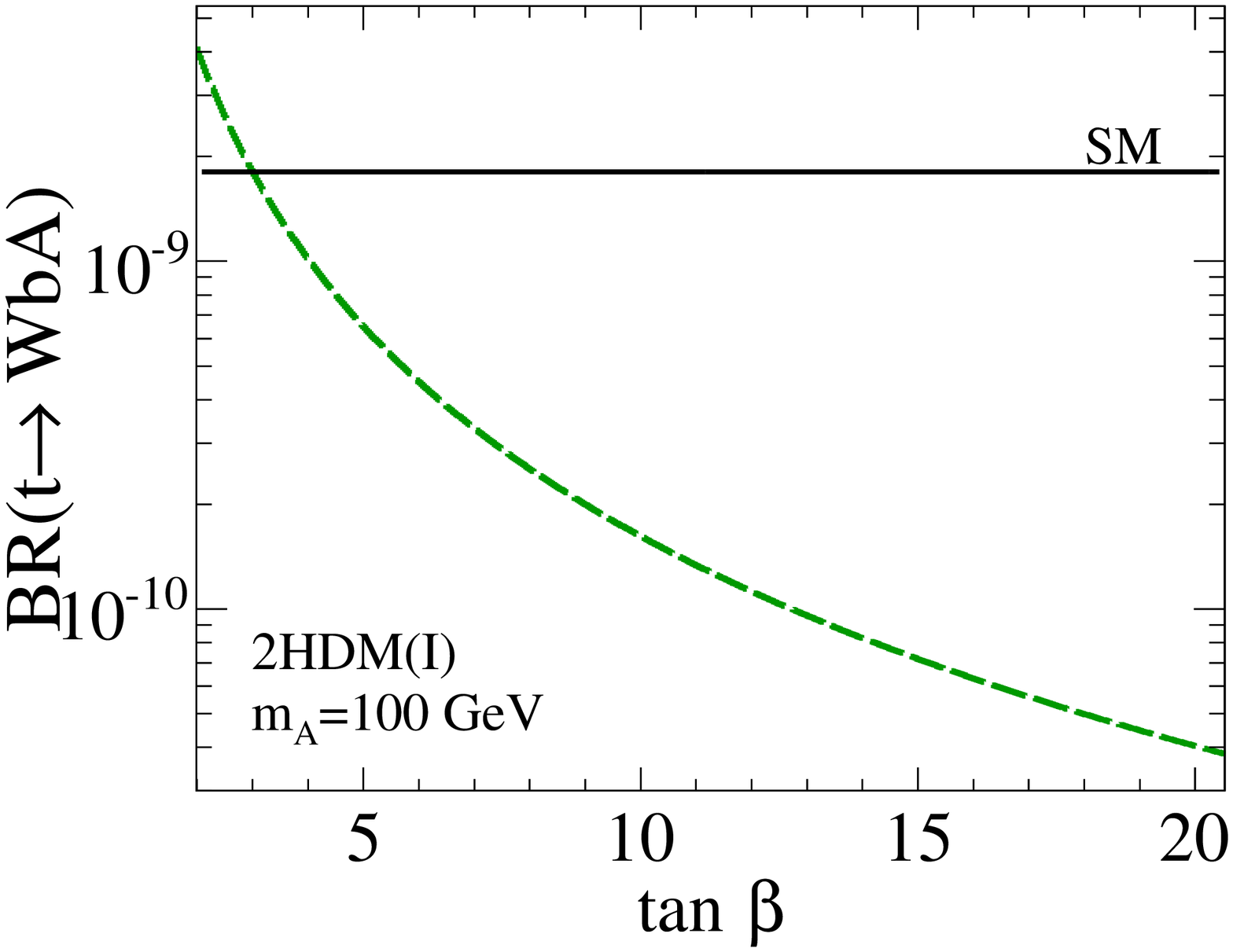}
	 \label{2hdmIA0vstB.FIG}
}
\subfigure[]{
	\includegraphics[width=0.47\textwidth]{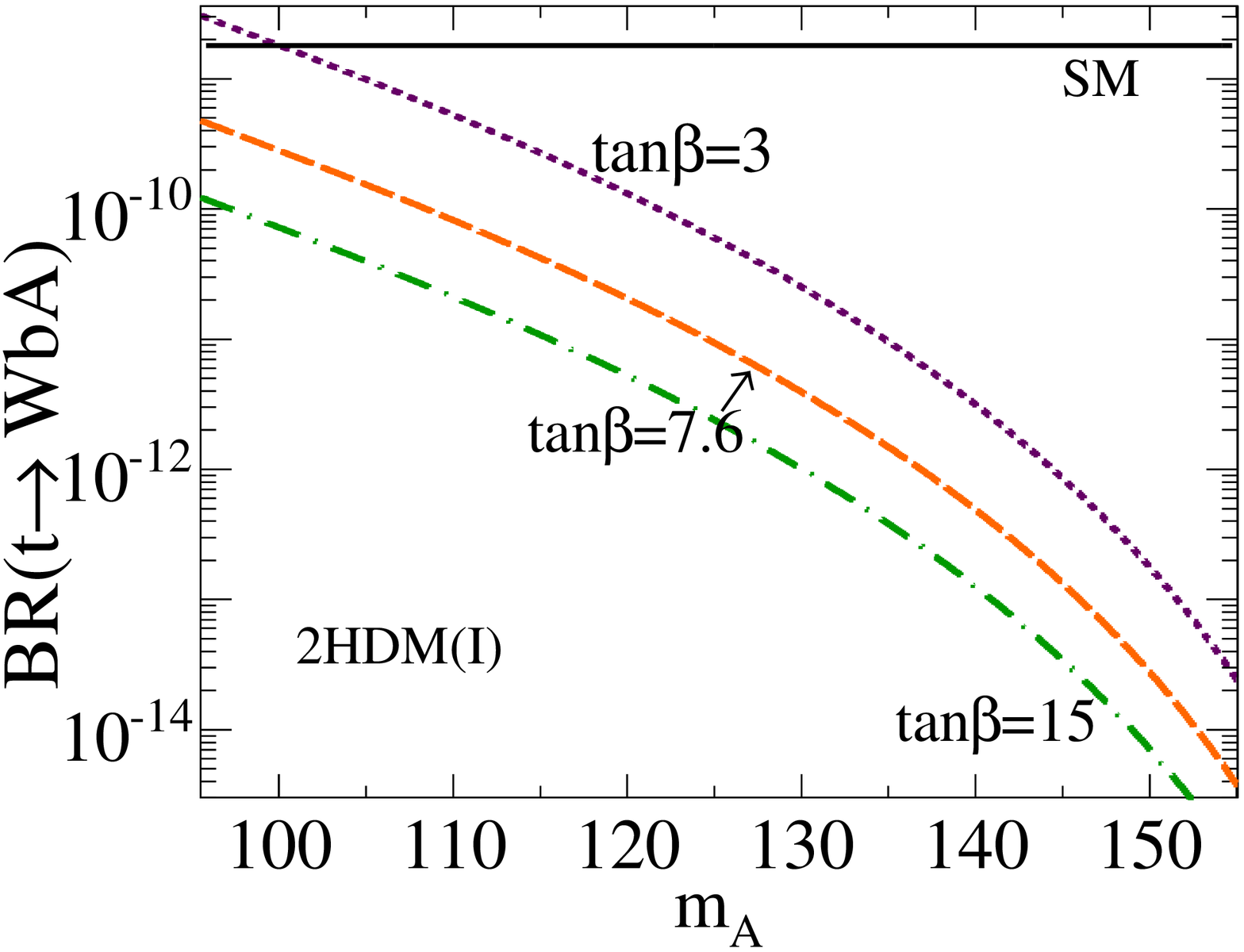}
	\label{2hdmIA0vsMA.FIG}
}
\caption{
The 2HDM(I) $\BR(t\rightarrow WbA)$ as a function of 
(a) $\tan\beta$ and (b) $m_{A}$ for 
$\tan\beta=3,~7.6,~15$ (short dash, long dash, dash-dot).
The solid line denotes the SM prediction, Eq.~(\ref{smBR.EQ}).
}
\label{2hdmI_A0.FIG}
\end{figure}

The decay rates for $t\rightarrow W^{*}bh$ and $t\rightarrow W^{*}bH$ as a function of (a) $\tan\beta$ and (b) $\Delta_{V}$ are shown in Fig.~\ref{2hdmI_HvsInput.FIG}.
Except for low value of $\tan\beta <3$, the rates are always smaller than the SM rate.
Beyond $\tan\beta\approx3$, the SM-like CP-even Higgs rates become independent of $\tan\beta$ and converge to asymptotic values;
for the non-SM-like Higgses, this occurs at $\tan\beta\approx15$. 
To see how this happens, note that the Yukawa couplings in the 2HDM(I) (Table \ref{2hdmICouple.TB}) take the simple form
\begin{equation}
 c_{1} \cot\beta + c_{2},
\end{equation}
where $c_{1,2}$ are elementary functions of $\Delta_{V}$, as seen in Table~\ref{2hdmICouple.TB}.
In the large $\tan\beta$ limit, the $c_{1}$ part vanishes, leaving the asymptotic value $c_{2}$.
In the SM-like limit, the $c_{2}$ terms are larger than the $c_{1}$ contributions, whereas the reverse holds in the non-SM-like limit.
We extract asymptotic values by observing that for a given CP-even Higgs the $c_{2}$ terms and $WWH$ couplings are the identical.
Consequently,
\begin{eqnarray}
 \underset{\tan\beta\rightarrow\infty}{\lim} \BR^{2HDM(I)}(t\rightarrow Wbh) &=& (1-\Delta_{V})^{2}\BR^{SM}(t\rightarrow Wbh)\nonumber\\
									    &=& \sin^{2}\left(\beta-\alpha\right)\BR^{SM}(t\rightarrow Wbh)
  \label{asympI.EQ}\\
 \underset{\tan\beta\rightarrow\infty}{\lim} \BR^{2HDM(I)}(t\rightarrow WbH) &=& (2\Delta_{V}-\Delta_{V}^{2})\BR^{SM}(t\rightarrow Wbh)\nonumber\\
									    &=& \cos^{2}\left(\beta-\alpha\right)\BR^{SM}(t\rightarrow Wbh).
 \label{asympII.EQ}
 \end{eqnarray}
For our choices of $\Delta_{V}$, the asymptotic rates in Fig.~\ref{2hdmI_brVstB.FIG} are 
\begin{eqnarray}
\underset{\tan\beta\rightarrow\infty}{\lim} \BR^{2HDM(I)}_{\Delta_{V}=0.7}(t\rightarrow WbH) &=& 0.910 \times \BR^{SM}(t\rightarrow Wbh),
 \label{smlikeH1Rate.EQ}\\
\underset{\tan\beta\rightarrow\infty}{\lim} \BR^{2HDM(I)}_{\Delta_{V}=0.05}(t\rightarrow Wbh) &=& 0.903 \times \BR^{SM}(t\rightarrow Wbh),
 \label{smlikeH2Rate.EQ}\\
\underset{\tan\beta\rightarrow\infty}{\lim} \BR^{2HDM(I)}_{\Delta_{V}=0.05}(t\rightarrow WbH) &=& 0.098 \times \BR^{SM}(t\rightarrow Wbh),
 \label{smunlikeH1Rate.EQ}\\
\underset{\tan\beta\rightarrow\infty}{\lim} \BR^{2HDM(I)}_{\Delta_{V}=0.7}(t\rightarrow Wbh) &=& 0.090 \times \BR^{SM}(t\rightarrow Wbh),
 \label{smunlikeH2Rate.EQ}
\end{eqnarray}
and agree well with numerical calculations.

The $\Delta_{V}$ dependence in Fig~\ref{2hdmI_brVsDV.FIG} and the relationship between $h$ and $H$ 
is indicative of much broader behavior found in all 2HDM variants.
To saturate the sum rule for the electroweak symmetry breaking \cite{Gunion:1989we}, the $hWW$ coupling ($g_{hWW}$) and the $HWW$ coupling ($g_{HWW}$) obey 
\begin{equation}
  g^{2}_{hWW}+g^{2}_{HWW}=g^{2}_{h^{SM}WW},
 \label{2hdmSumRule.EQ}
\end{equation}
where $g_{h^{SM}WW}$ is the SM $hWW$ coupling. 
For $h$ and $H$ with degenerate masses, 
\begin{equation}
  BR(t\rightarrow Wbh) + BR(t\rightarrow WbH) = \BR^{SM}(t\rightarrow Wbh)+\mathcal{O}(\cot^{2}\beta).
 \label{2hdmSumRule2.EQ}
\end{equation}
Indeed, Eqs.~(\ref{asympI.EQ}) and (\ref{asympII.EQ}) satisfy this relationship. 
Furthermore, this can be extended to an arbitrary number of scalar $SU(2)_{L}$ doublets and singlets~\cite{Gunion:1989we}.
Though mass splittings, etc., will break this equality, it provides a useful estimate for processes involving transitions in models with additional scalar $SU(2)_{L}$ doublets and singlets.

\subsubsection{BR$(t\rightarrow Wbh/H)$ vs $m_{h/H}$}
As a function of mass, we plot in Fig.~\ref{2hdmI_HvsMH.FIG} the decay rates for $t\rightarrow W^{*}bH$ where $H$ is a non-SM-like CP-even Higgs;
the mass of the SM-like Higgs is taken to be $125.5$ GeV.
For a mass below (above) $110$ GeV, we observe that transition rate to a non-SM-like Higgs remains above (below) the SM rate.
As the scalar mass decreases and the $W^{*}$ comes closer to being on-shell, 
the availability of phase space greatly ameliorates the coupling suppression associated with $\Delta_{V}$.
However, despite this relief, the transition rate to a non-SM-like $H$ stays below the SM rate for much of the parameter space.
The insensitivity to large and moderate $\tan\beta$ seen in Fig.~\ref{2hdmI_HvsMH.FIG} is consistent with previous discussions.

\begin{table}
\caption{$\BR(t\rightarrow WbH)$ for Benchmark Values of Higgses in the 2HDM(II)}
 \begin{center}
\begin{tabular}{|c|c|c|c|}
\hline 
 $H~(125.5~{\rm GeV})$ &  $\Delta_{V}$ & $\tan\beta$ & $\BR^{2HDM(II)}(t\rightarrow WbH)$  \tabularnewline\hline\hline 
 $h$ &  $5\times10^{-5}$ & $3$   & $1.813 \times 10^{-9}$     \tabularnewline\hline
 $h$ &                   & $7.6$ & $1.809 \times 10^{-9}$     \tabularnewline\hline
 $H$ &  $0.99$           & $3$   & $1.798 \times 10^{-9}$     \tabularnewline\hline 
 $H$ &                   & $7.6$ & $1.802 \times 10^{-9}$     \tabularnewline\hline
 $h$ &  $0.99$           & $3$   & $1.440 \times 10^{-10}$    \tabularnewline\hline
 $h$ &                   & $7.6$ & $4.990 \times 10^{-11}$    \tabularnewline\hline
 $A~(100~{\rm GeV})$ &   & $3$   & $1.760 \times 10^{-9}$    \tabularnewline\hline
 $A~(100~{\rm GeV})$ &   & $7.6$ & $1.007 \times 10^{-9}$  \tabularnewline\hline\hline
\end{tabular}
\label{2hdmIIBench.TB}
\end{center}
\end{table}

\begin{figure}[tb]
\centering
\subfigure[]{
	\includegraphics[width=0.47\textwidth]{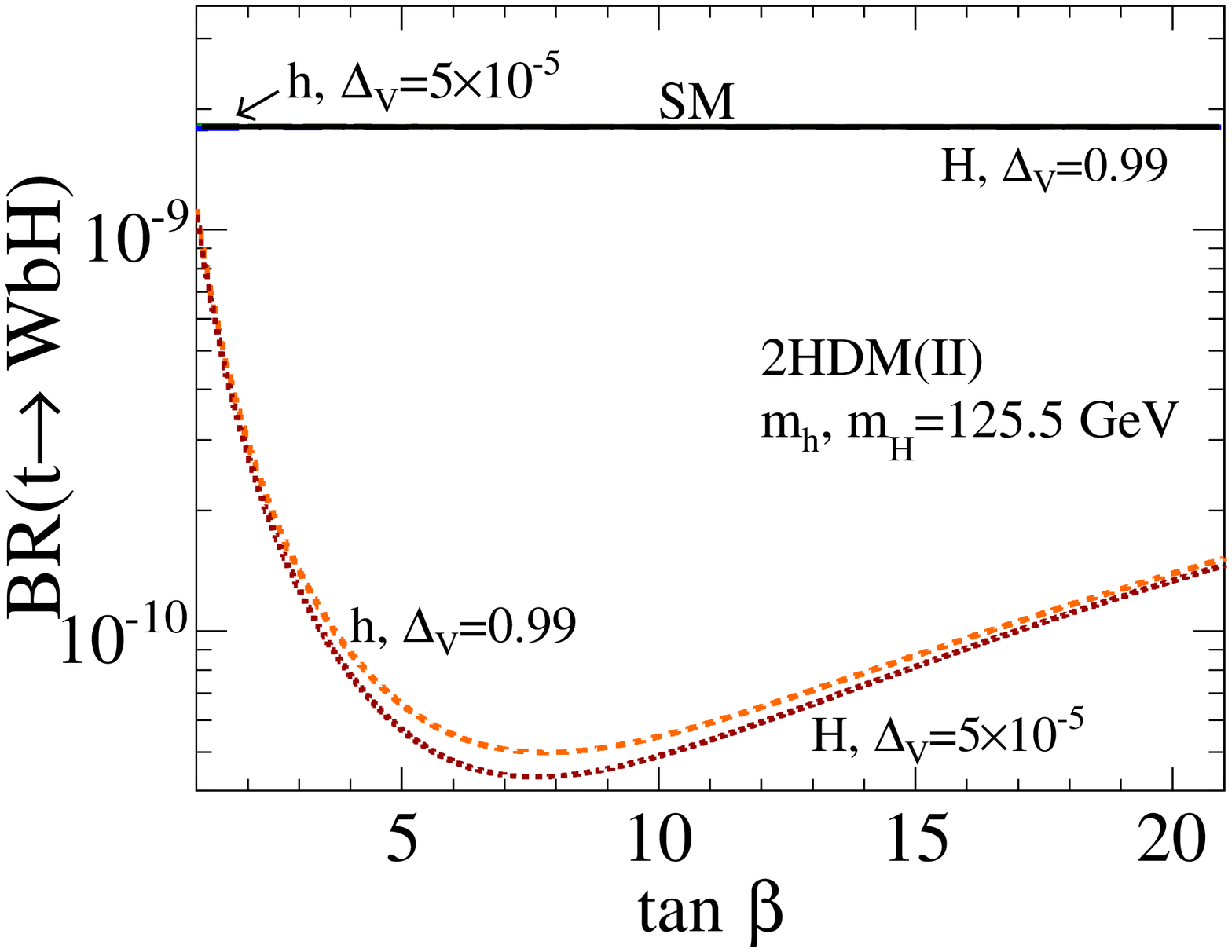}	
	\label{brVstB.FIG}
}
\subfigure[]{
	\includegraphics[width=0.47\textwidth]{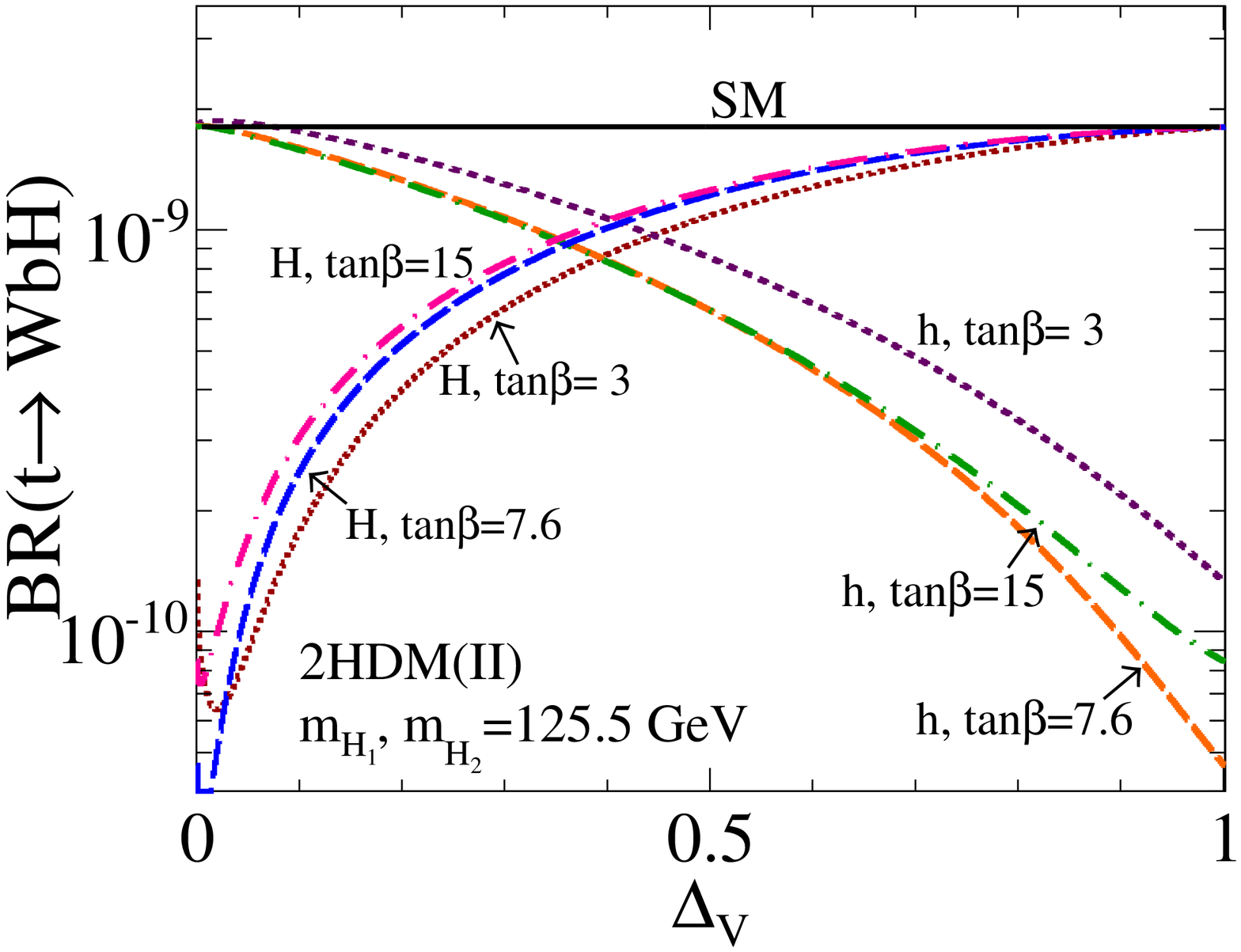}	
	\label{brVsDV.FIG}
}
\caption{
The 2HDM(II) $\BR(t\rightarrow WbH)$ as a function of 
(a) $\tan\beta$ for SM-like $h$ (long dash), $H$ (dash-dot), and non-SM-like $h$ (short dash), $H$ (dot);
(b) $\Delta_{V}$ for $h$ at $\tan\beta = 3~7.6,~15$ (short dash, long dash, dash-dot), and for $H$ (dot, long dash, dash-dot).
The solid line denotes the SM prediction, Eq.~(\ref{smBR.EQ}).
}
\label{HvsInput.FIG}
\end{figure}	

\subsubsection{BR$(t\rightarrow WbA)$ vs $\tan\beta$, $m_{A}$}
Here, we consider the decay rate to the CP-odd Higgs, $t\rightarrow W^{*}bA$.
Fig. \ref{2hdmI_A0.FIG} shows $\BR(t\rightarrow WbA)$ as a function of (a) $\tan\beta$ and (b) $m_{A}$.
Except for very low $\tan\beta$ and mass, the branching fraction remains well below the SM prediction for much of the parameter space,
approximately equaling it at $\tan\beta\simeq3$ for $m_{A}=100$ GeV.
Due to CP-invariance in the gauge sector there is no tree-level $AWW$ contribution.
And since the $f\overline{f}A$ couplings are independent of $\Delta_{V}$, the decay rate is fixed entirely by $m_{A}$ and $\tan\beta$. 
Destructive interference still exists, however, since the $t\overline{t}A$ and $b\overline{b}A$ vertices differ by a minus sign.
A quadratic dependence on $\cot\beta$ is the consequence the $f\overline{f}A$ coupling $(\propto \cot\beta)$.
See Table~\ref{2hdmICouple.TB}.
Despite this monotonic dependence on $\tan\beta$, which implies that $\BR(t\rightarrow WbA)$ is a direct measure of $\tan\beta$ were it to be measured, 
the recuperation of available phase space at low $m_{A}$ is unable to compensate for the $\cot^{2}\beta$ suppression.


\subsection{Type II 2HDM BR$(t\rightarrow WbH)$}
\label{2hdmIIBR.SEC}

\begin{figure}[tb]
\centering
\subfigure[]{
	\includegraphics[width=0.47\textwidth]{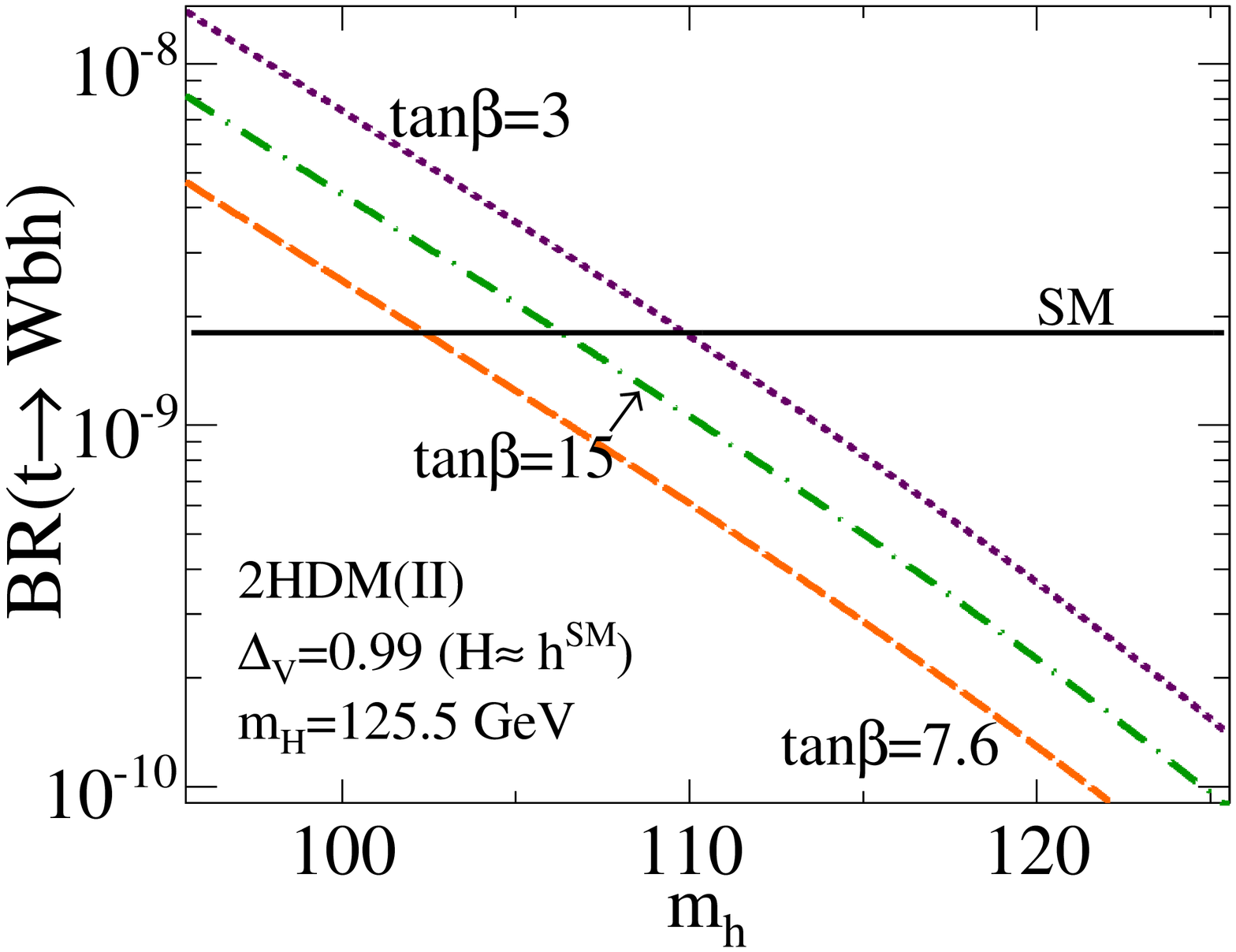}
}
\subfigure[]{
	\includegraphics[width=0.47\textwidth]{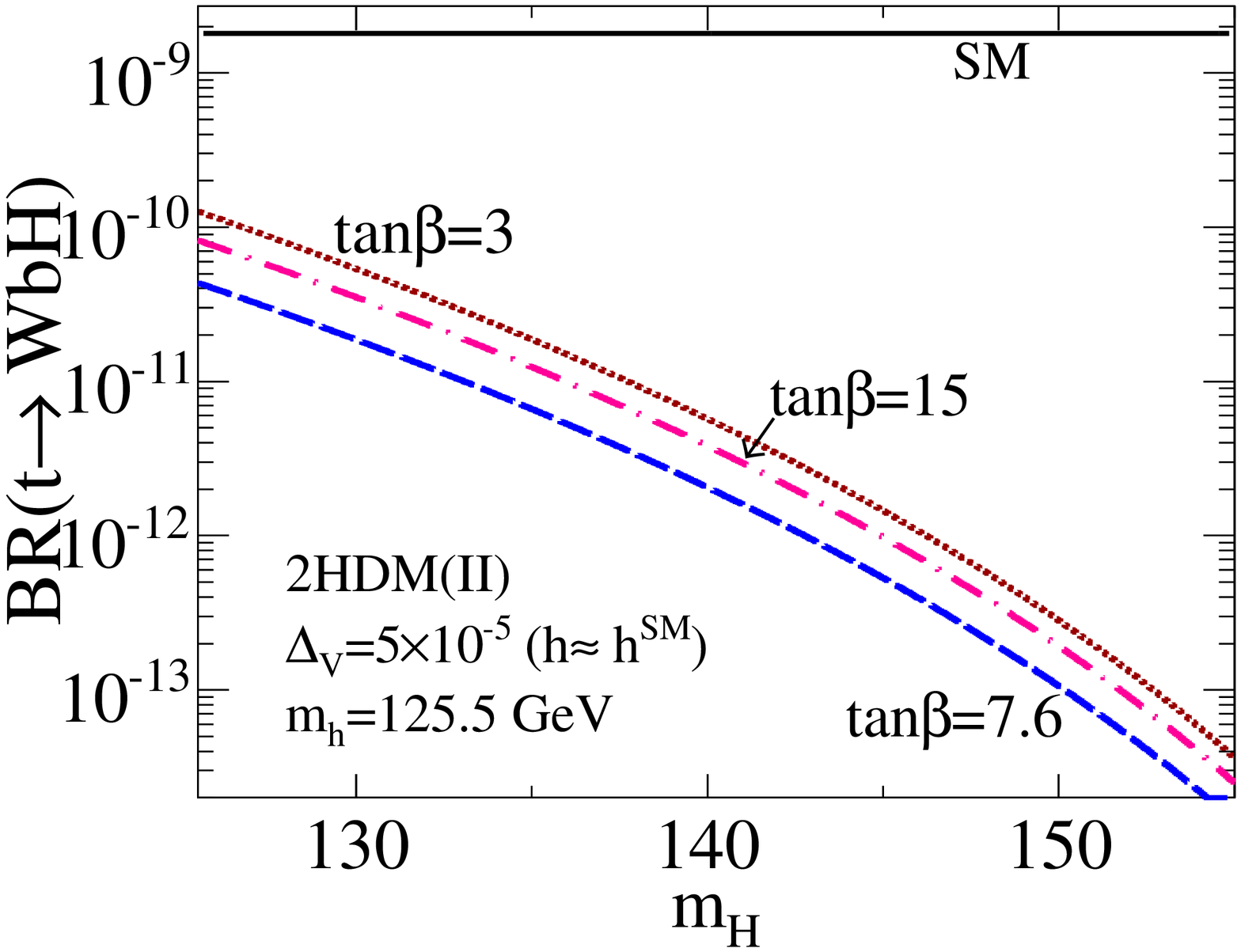}	
}
\caption{
The 2HDM(I) $\BR(t\rightarrow WbH)$ as a function of mass for a non-SM-like 
(a) $h$ and (b) $H$ assuming $\tan\beta=3,~7.6,~15$ (short dash, long dash, dash-dot).
The solid line denotes the SM prediction, Eq.~(\ref{smBR.EQ}).
}
\label{HvsMH.FIG}
\end{figure}

We report here the behavior of $\BR(t\rightarrow WbH)$, where $H$ represents $h,~H,$ or $A$ in the 2HDM(II). 
The same values of $\tan\beta$ are used here as the Type I case.
To avoid constraints, we choose a $\Delta_{V}$ that corresponds to a light Higgs with $\cos(\beta-\alpha)=0.01$, i.e.,
\begin{equation}
  \Delta_{V}=5.\times10^{-5}~(0.99)\quad\text{for}\quad h~(H)\approx h^{SM}.
\end{equation}
Table~\ref{2hdmIIBench.TB} lists values of the branching fraction for several Higgses and benchmark parameter values.
In the following figures, the predicted SM decay rate is shown as a black, solid line labeled by ``SM''.

\subsubsection{BR$(t\rightarrow Wbh, H)$ vs $\tan\beta$, $\Delta_{V}$}

Figure~\ref{HvsInput.FIG} depicts the branching ratio $\BR(t\rightarrow WbH)$ for both of the CP-even Higgses as a function of 
(a) $\tan\beta$ for SM-like and non-SM-like $h$ and $H$, and (b) $\Delta_{V}$ for small and large values of $\tan\beta$.

In Fig~\ref{brVstB.FIG}, for SM-like Higgses, the branching fraction is indistinguishable from the SM prediction as a function of $\tan\beta$;
for non-SM-like Higgses, however, the rates minimize for $\tan\beta=7\sim8$.
This dependence on $\tan\beta$ is indicative of a playoff between the $t\overline{t}H$ and the $b\overline{b}H$ couplings.
In the SM limit, this specific behavior is suppressed for SM-like Higgs bosons because the couplings to these bosons grow independent of $\tan\beta$. 
When $h$ is non-SM-like $(\Delta_{V}=0.99)$, sensitivity to $\tan\beta$ maximizes because the $\tan\beta$-independent parts of the fermionic Higgs couplings nearly cancel.
As $\tan\beta$ grows, the contribution from $t\overline{t}h~(\propto \cot\beta)$ 
runs $\BR(t\rightarrow Wbh)$ down until the $b\overline{b}h$ contribution $(\propto \tan\beta)$ takes over at $\tan\beta\approx 7.6$.
When $H$ is non-SM-like $(\Delta_{V}=5.\times 10^{-5})$ we expect and observe similar behavior as the non-SM-like $h$ case.

Much of the relationship between $h$ and $H$ observed in in Fig.~\ref{brVsDV.FIG} is type-independent and the discussion can be found in the Type I scenario.
For a light Higgs, we indeed see that at $\tan\beta=7.6$ transition rates are minimized for all values of $\Delta_{V}$.
For a heavy Higgs, however, this value of $\tan\beta$ only minimizes the rate in the $h\rightarrow h^{SM}$ limit, 
in which case the $t\rightarrow H$ transition rate vanishes.
The $t\rightarrow H$ rate minimum occurs at larger $\Delta_{V}$ with decreasing $\tan\beta$ because the $t\overline{t}H~(b\overline{b}H)$  contribution becomes numerically larger (smaller).


\subsubsection{BR$(t\rightarrow Wbh, H)$ vs $m_{H}$}

Figure~\ref{HvsMH.FIG} presents the $t\rightarrow W^{*}bH$ branching ratio for a non-SM-like Higgs boson as a function of mass.
For the mass window given in Eq.~(\ref{mAMass.EQ}), we find considerable enhancement in the decay rate relative to the SM rate due to the increase in available phase space, overcoming the coupling suppression associated with scalars that have non-SM-like coupling.

\subsubsection{BR$(t\rightarrow WbA)$ vs $\tan\beta$, $m_{A}$}
\begin{figure}[tb]
\centering
\subfigure[]{
	\includegraphics[width=0.47\textwidth]{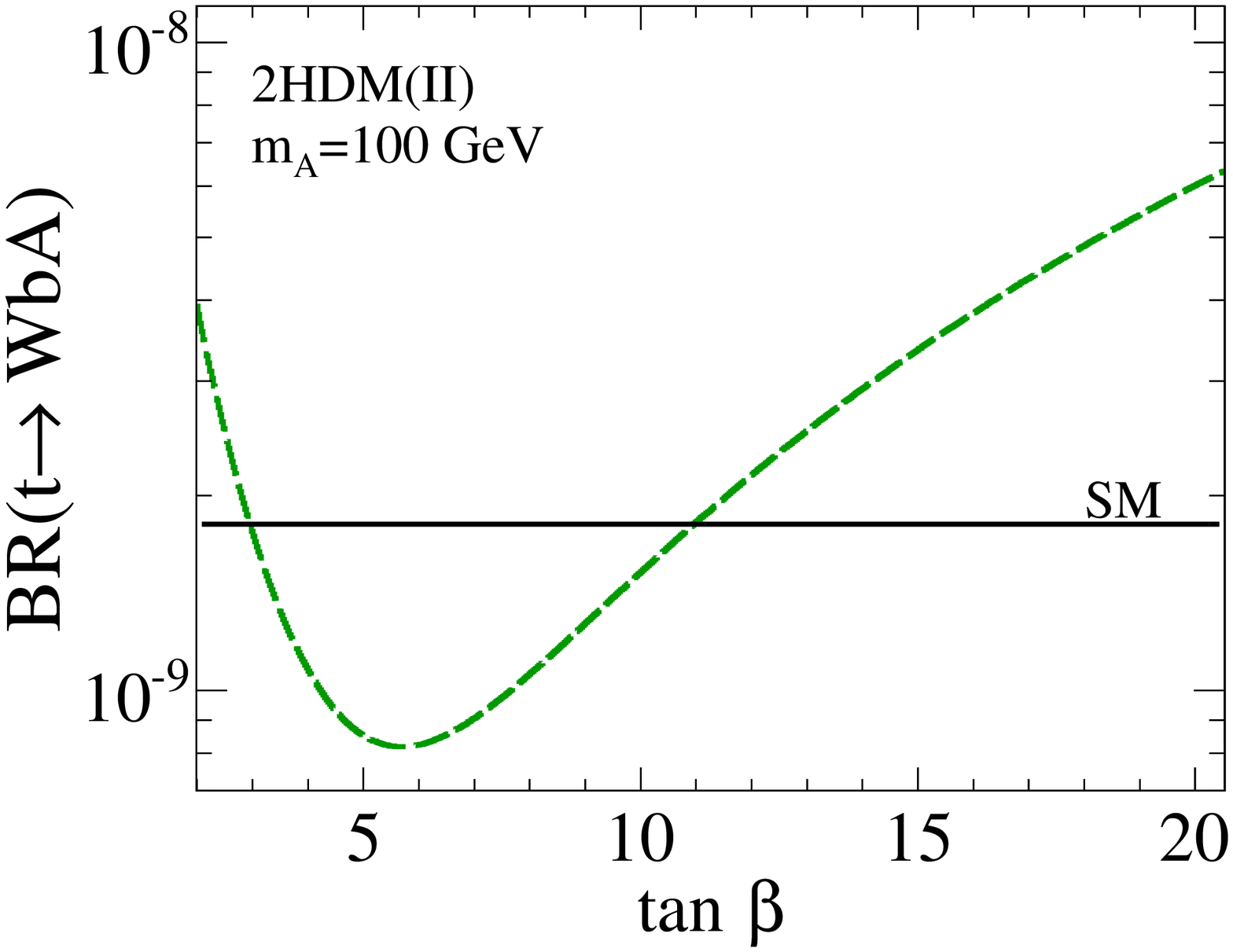}
	 \label{A0vstB.FIG}
}
\subfigure[]{
	\includegraphics[width=0.47\textwidth]{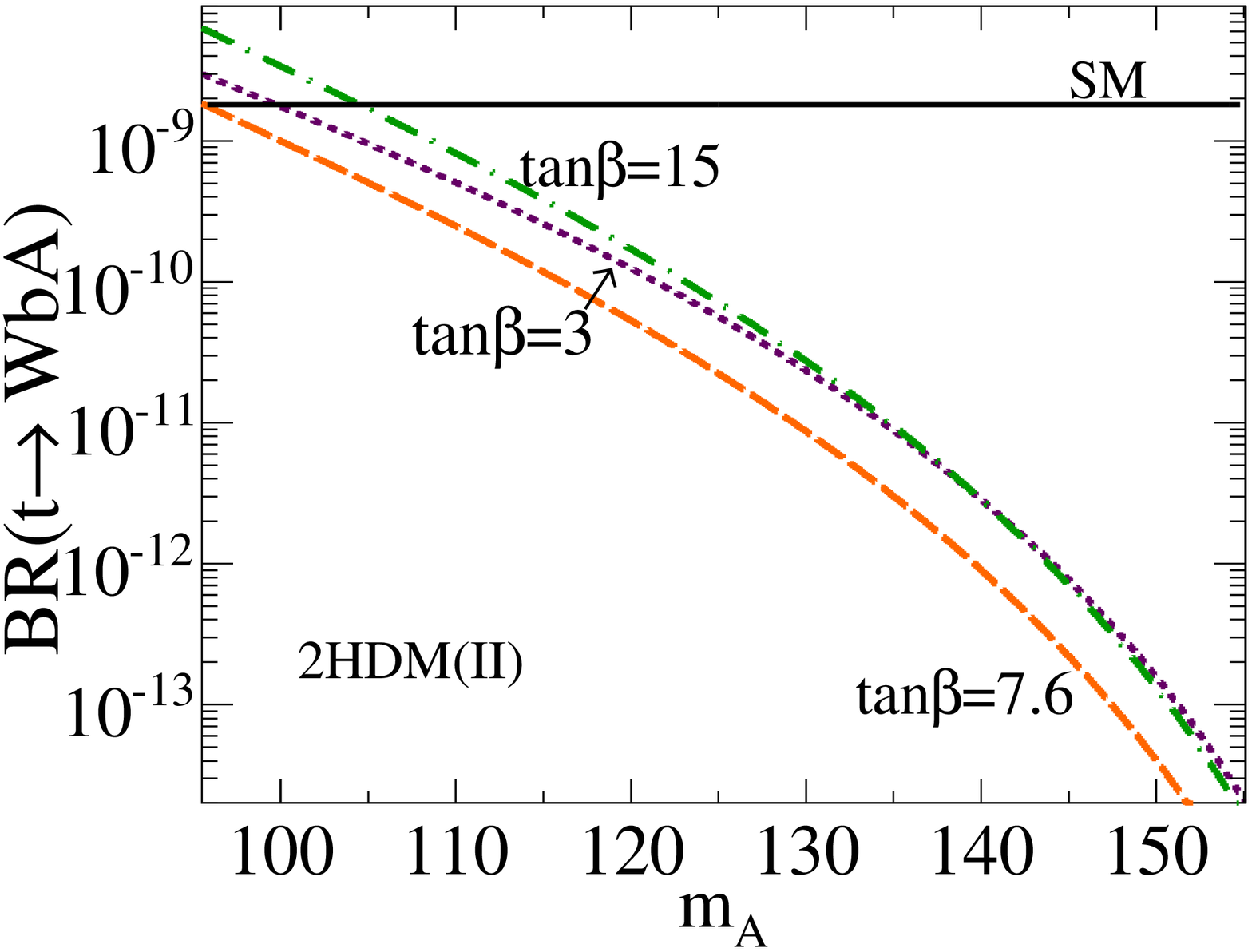}
	\label{A0vsMA.FIG}
	}
\caption{
The 2HDM(II) $\BR(t\rightarrow WbA)$ as a function of 
(a) $\tan\beta$ and (b) $m_{A}$ for $\tan\beta=3,~7.6,~15$ (short dash, long dash, dash-dot).
The solid line denotes the SM prediction, Eq.~(\ref{smBR.EQ}).
}
\label{A0vsX.FIG}
\end{figure}

Turning to the CP-odd Higgs decay channel, $t\rightarrow W^{*}bA$,
we note that many of the arguments made in the 2HDM(I) case carry over to this situation. 
Unlike the Type I scenario, however, there is only {\it constructive} interference between the fermion contributions.
Figure~\ref{A0vsX.FIG} shows $\BR(t\rightarrow WbA)$ as a function of (a) $\tan\beta$ and (b) $m_{A}$.

In Fig.~\ref{A0vstB.FIG}, due to an accidental cancellation, the branching fraction minimizes at $\tan\beta\approx 5.8$,
which is unsurprisingly close to the $t\rightarrow H^{+}b$ minimum at $\tan\beta=\sqrt{m_{t}/m_{b}}\approx 7.6$.
At $\tan\beta\approx 5.8$, 
the $t\overline{t}A$ coupling $(\propto\cot\beta)$ and the $b\overline{b}A$ coupling $(\propto\tan\beta)$ contribute equally.  
At smaller values of $\tan\beta$, $t\overline{t}A$ is the dominant term but is driven down by an increasing $\tan\beta$; 
and at larger values, $b\overline{b}A$ is the dominant term, which ramps up the rate.
In the large $\tan\beta$ limit, the $t\overline{t}A$ graph becomes negligible and the rate becomes quadratically with $\tan\beta$.

In Fig.~\ref{A0vsMA.FIG}, we observe a similarity between $A$ and the non-SM-like Higgs boson, $H_{X}$. 
We attribute this to a similarity of contributing diagrams. 
For example: the $WWA$ vertex does not exist because of CP-invariance, and by virtue of being {\it non}-SM-like, the $WWH_{X}$ vertex is considerably suppressed. 
In this domain, fermionic couplings to $A$ and $H_{X}$ also have the same dependence on $\tan\beta$.

\section{OBSERVATION PROSPECTS AT COLLIDERS}
\label{LHC.SEC}
In this section, we estimate observation prospects at current and future colliders.
The 14 TeV LHC $t\overline{t}$ production cross section at NNLO in QCD has been calculated~\cite{Czakon:2013goa} to be
\begin{equation}
\sigma_{LHC14}^{NNLO}(t\overline{t})= 933~\text{pb}.
\end{equation}
The SM $pp\rightarrow t\overline{t}\rightarrow WW^{*}b\overline{b}h$ cross section at the LHC is thus estimated to be
\begin{equation}
 \sigma_{LHC14}(pp\rightarrow t\overline{t}\rightarrow WW^{*}b\overline{b}h)
 \approx2\times\sigma^{NNLO}_{LHC14}(t\overline{t})\times BR(t\rightarrow Wbh) 
 = {3.4}~\text{ab}.
\label{lhcppWbh.EQ}
 \end{equation}
The factor of two in the preceding line accounts for either top or antitop quark decaying into the Higgs.
To assure a clear trigger and to discriminate against the large SM backgrounds, we require at least one $W$ boson decaying leptonically ($\ell = e, \mu$), i.e.
\begin{equation}
 \BR(WW^{*}\rightarrow \ell^{+}\ell^{'-}\nu_{\ell}\overline{\nu}_{\ell'} + jj\ell^{\pm}\overset{(-)}{\nu_{\ell}})\approx 0.33.
\end{equation}
The total cross section for an arbitrarily decaying $h$ is therefore estimated to be 
\begin{eqnarray}
 \sigma_{LHC14}(pp\rightarrow t\overline{t}\rightarrow WW^{*}b\overline{b}h\rightarrow 
 h(\ell^{+}\ell^{'-}\nu_{\ell}\overline{\nu}_{\ell'} + jj\ell^{\pm}\overset{(-)}{\nu_{\ell}})) &\approx&  1.1~\text{ab}.
\label{lhcppWhbDecay.EQ}
 \end{eqnarray}
Higgs branching fractions and detector efficiencies will further suppress this rate.
Such a small cross section means that observing this SM process will be challenging. 
Following the same procedure, we estimate Eq.~(\ref{lhcppWhbDecay.EQ}) for several proposed colliders and collider upgrades; 
the results are given in Table~\ref{collider.TB}.

\begin{table}
\caption{Cross sections for $t\overline{t}$ and $t\overline{t}\rightarrow WW^{*}b\overline{b}h\rightarrow 
 h(\ell^{+}\ell^{'-}\nu_{\ell}\overline{\nu}_{\ell'} + jj\ell^{\pm}\overset{(-)}{\nu_{\ell}})$ at 
14~\cite{Czakon:2013goa}, 33~\cite{ttbar33TeV:2012}, and 100~\cite{Baur:2002ka} TeV $pp$, and 350 GeV $e^{+}e^{-}$~\cite{seidelTop:2012} Colliders.}
 \begin{center}
\begin{tabular}{|c|c|c|c|c|}
\hline 
Process  & 14 TeV $pp$ & 33 TeV $pp$ & 100 TeV $pp$ & 350 GeV $e^{+}e^{-}$  \tabularnewline
\hline\hline
$\sigma(t\overline{t})$[pb]  & $933$ & $5410$ & $2.7\times10^{4}$  & $0.45$  \tabularnewline \hline
$\sigma(t\overline{t}\rightarrow 
 h(\ell^{+}\ell^{'-}\nu_{\ell}\overline{\nu}_{\ell'} + jj\ell^{\pm}\overset{(-)}{\nu_{\ell}}))$ [ab] & $1.1$ & $6.5$  & $32$ & $5\times10^{-4}$      \tabularnewline \hline
\hline
\end{tabular}
\label{collider.TB}
\end{center}
\end{table}

\section{SUMMARY AND CONCLUSION}
\label{conc.SEC}

Given the discovery of a SM-like Higgs boson, we have recalculated the rare top quark decay mode $t\rightarrow W^{*}bh$,  
where $h$ represents the SM Higgs boson. 
We have extended this calculation to include the effects of anomalous $t\overline{t}h$ couplings originating from effective operators 
as well as both CP-even and the single CP-odd scalars in the CP-conserving 2HDM Types I and II. 
The most updated model constraints have been reported. 
We summarize our results:
\begin{enumerate}[(i)]

 \item The SM predicts a $t\rightarrow W^{*}bh$ branching ratio of
 \begin{equation}
 \BR(t\rightarrow Wbh) = 1.80\times 10^{-9}\quad\text{for}\quad m_{h}=125.5~\text{GeV}.
 \label{conBR.EQ}
 \end{equation}
This is the leading $t\rightarrow h$ transition, five orders of magnitude larger than the next channel $t\rightarrow ch$.  See Eq.~(\ref{smBR.EQ}).

 \item Present LHC Higgs constraints on anomalous $t\overline{t}h$ couplings permit up to a factor of two enhancement of the $t\rightarrow W^{*}bh$ transition. See Eq.~(\ref{eftBRLimitI.EQ}).

 \item The operator $\overline{\mathcal{O}}_{t2}$, 
 which selects different kinematic features than either $\mathcal{O}_{t1}$ or $\overline{\mathcal{O}}_{t1}$,
 results in comparable enhancement of the $t\rightarrow W^{*}bh$ transition. See Fig.~\ref{effBR.FIG}.
 
  \item In the 2HDM(I), decays to CP-even Higgses do not decouple in the large $\tan\beta$ limit and their rates approach asymptotic values that are functions of the anomalous $WWh$ coupling. They are given in Eqs.~(\ref{asympI.EQ}) and (\ref{asympII.EQ}).

 \item In the Type I (II) 2HDM, due to the increase in available phase space, the branching ratio to a light, non-SM-like Higgs boson can as much as $2~(7)$ times larger than Eq.~(\ref{conBR.EQ}). 

 \item In the Type I (II) 2HDM, the branching ratio to a light, CP-odd Higgs can be as much as $1.6~(3)$ times larger than Eq.~(\ref{conBR.EQ}). 
 
 \item The $pp\rightarrow t\overline{t}\rightarrow WW^{*}b\overline{b}h\rightarrow 
 h(\ell^{+}\ell^{'-}\nu_{\ell}\overline{\nu}_{\ell'} + jj\ell^{\pm}\overset{(-)}{\nu_{\ell}})$ production cross section at the 14 TeV LHC and future colliders have been estimated [Eq.~(\ref{lhcppWhbDecay.EQ})]; a few $t\rightarrow W^{*}bh$ events over the full LHC lifetime. 
      Due to enhancements in gluon distribution functions, any increase in collision energies can greatly increase this rate.

\end{enumerate}

\vspace*{0.3cm}

\noindent {\it Acknowledgments:} 
We wish to thank Martin Einhorn and Jos\'e Wudka for their constructive and insightful comments.
We also wish to thank Neil Christensen and Joshua Sayre for useful comments and discussions, especially pertaining to the workings of CalcHEP.
R.R.~would like to thank the Particle Theory and Cosmology Group at Tohoku University for their warm hospitality while this manuscript was written. The work of T.H.~was supported in part by the U.S.~Department of Energy under Grant No. DE-FG02-95ER40896 and in part by the PITT-PACC.
R.R.~acknowledges support from the University of Pittsburgh.


\pagebreak


\end{document}